\documentclass{elsartModified}

\usepackage{verbatim}
\usepackage{amsmath}
\usepackage{latexsym}
\usepackage{listings}
\usepackage{ifthen}
\usepackage{enumerate}
\usepackage[normalem]{ulem}
\usepackage{epsfig}
\usepackage[bookmarks=false, pdfauthor={}, pdftitle={}]{hyperref} 
\usepackage{boxedminipage}

\newcommand{\lstbox}[1]{\mbox{\lstinline{#1}}}
\newcommand{\mathbox}[1]{\mbox{${#1}$}}

\newcommand{\figskip}{\medskip}
\newcommand{\relatedWorkParagraph}[1]{%
\vspace{-88\in}
\paragraph*{#1}}
\newcommand{\vspaceSection}{\vspace{-92\in}}
\newcommand{\introSection}[1]{
\vspace{-22\in}
\subsection*{#1}
\vspaceSection}
\newcommand{\mySection}[1]{
\vspace{-22\in}
\section{#1}
\vspaceSection}
\newcommand{\mySubsection}[1]{
\vspace{-22\in}
\subsection{#1}
\vspaceSection}
\newcommand{\mySubsubsection}[1]{
\vspace{-22\in}
\subsubsection{#1}
\vspace{-22\in}}
\newcommand{\citeGFP}{\cite{JJ97,Hinze00-POPL,RodriguezJJGKO08}}
\newcommand{\citeAP}{\cite{PalsbergPL97-SCP,LieberherrPO04,LaemmelVV03-AOSD}}
\newtheorem{advice}{Advice}
\newcommand{\hsnote}[1]{\footnote{\normalfont\emph{Note on Haskell}: #1}}
\newlength{\basewidth}
\newlength{\topwidth}
\newcommand{\superimpose}[2]{%
  \settowidth{\basewidth}{#1}%
  \settowidth{\topwidth}{#2}%
  \ifthenelse{\lengthtest{\basewidth > \topwidth}}%
    {\makebox[0pt][l]{#1}\makebox[\basewidth]{#2}}%
    {\makebox[0pt][l]{#2}\makebox[\topwidth]{#1}}%
}
\newcommand{\hs}[1]{\lstinline[columns=fullflexible]{#1}}
\newcommand{\true}{\ensuremath{\mathsf{True}}}
\newcommand{\false}{\ensuremath{\mathsf{False}}}
\newcommand{\smbox}[1]{\mbox{{\footnotesize #1}}}
\newcommand{\rewritesto}{\ensuremath{\to}}
\newcommand{\idT}{\ensuremath{\w{id}}}
\newcommand{\failT}{\ensuremath{\w{fail}}}
\newcommand{\failure}{\ensuremath{{\uparrow}}}
\newcommand{\rec}[2]{\ensuremath{\mu{}#1.#2}}

\newcommand{\sequ}[2]{\ensuremath{#1;#2}}
\newcommand{\choicechar}{\ensuremath{\superimpose{$\leftarrow$}{$+$}}}
\newcommand{\choice}[2]{\ensuremath{#1\,\choicechar\,#2}}
\newcommand{\allchar}{\ensuremath{\Box}}
\newcommand{\all}[1]{\ensuremath{\allchar(#1)}}

\newcommand{\onechar}{\ensuremath{\Diamond}}
\newcommand{\one}[1]{\ensuremath{\onechar(#1)}}

\newcommand{\apply}[2]{\ensuremath{#1\,@\,#2}}

\newcommand{\cwtj}[3]{\ensuremath{#1{}\vdash{}#2:#3}}
\newcommand{\ioj}[2]{\ensuremath{#1 \leadsto #2}}

\newcommand{\ir}[3]{\ensuremath{%
\begin{array}[b]{c}
#2
\\ \hline
#3
\end{array}\hfill
\begin{tabular}[b]{r}\vspace{.5em}\tg{#1}
\end{tabular}
\medskip
}}
\newcommand{\tg}[1]{\ensuremath{{[}\mbox{{\ensuremath{\mathsf{#1}}}}{]}}}

\newcommand{\ax}[2]{\ensuremath{%
\begin{array}[b]{c}
#2
\end{array}\hfill
\begin{tabular}[c]{r}\vspace{.5em}{\tg{#1}}
\end{tabular}
\medskip
}}

\newcommand{\dnp}{\ \wedge\ }

\newcommand{\noskip}{\topsep0pt \parskip0pt \partopsep0pt}

\newcommand{\w}[1]{\ensuremath{\mathit{#1}}}

\lstdefinelanguage{haskell}{%
  numbers=none,
  morekeywords={do,module,qualified,if,then,else,import,let,case,of,in,class,instance,data,newtype,where,otherwise,deriving,type,extends,public,forall},
  sensitive=true,
  commentstyle=\rmfamily\upshape,
  morecomment=[l]{--},
  morecomment=[s]{/*}{*/},
  morestring=[b]",
}

\lstdefinelanguage{isabelle}{%
  numbers=none,
  morekeywords={inductive,intro,intros,back,text,primrec,defs,recdef,if,case,then,else,lemma,theorem,forall,apply,of,rule,done,datatype,types,axioms,consts},
  sensitive=true,
  morecomment=[s]{\{*}{*\}},
  literate={~}{{\mbox{$\neg$}}}1 {"}{}1 {\\<times>}{{\,\,$\times$\,\,}}1 {[|}{{\mbox{$[|$}}}1 {|]}{{\mbox{$|]$}}}1 {<=}{{\,$\leq$\,\,}}1 {==}{{\,\,$=$\,\,}}1 {\\<and>}{{\,$\wedge$\,\,\,}}1 {\\<or>}{{\,$\vee$\,\,\,}}1 {\\<exists>}{{$\exists$\,}}1 {\\<forall>}{{$\forall$\,}}1 {=>}{{\,\,$\to$\,\,}}1 {==>}{{\,\,$\Longrightarrow$\,\,}}1 {-->}{{\,\,$\Longrightarrow$\,\,}}1 {~=}{{\,$\not=$\,}}1 {?}{{$\exists$\,}}1 {\\<lambda>}{{$\lambda$\,}}1 {\\<Union>}{{$\bigcup$\,}}1 {\\<union>}{{$\cup$\,}}1
}

\begin{document}

\begin{frontmatter}

  \title{Programming errors in traversal programs over structured
    data}\thanks{The paper and accompanying source code are available online:
\begin{quote}
\begin{itemize}
\item \url{http://userpages.uni-koblenz.de/~laemmel/syb42}
\item \url{http://code.google.com/p/strafunski/}
\end{itemize}
\end{quote}}

\author{%
Ralf~L\"ammel{$^{1}$}
\and
Simon~Thompson{$^{2}$}
\and 
Markus Kaiser{$^{1}$}}

{\small\sl $^1$\ University of Koblenz-Landau, Germany}\\
{\small\sl $^{2}$\ University of Kent, UK}

\bigskip


\maketitle

\begin{abstract}
  Traversal strategies \`a la Stratego (also \`a la Strafunski and
  `Scrap Your Boilerplate') provide an exceptionally versatile and
  uniform means of querying and transforming deeply nested and
  heterogeneously structured data including terms in functional
  programming and rewriting, objects in OO programming, and XML
  documents in XML programming.

  \smallskip

  However, the resulting traversal programs are prone to programming
  errors. We are specifically concerned with errors that go beyond
  conservative type errors; examples we examine include divergent
  traversals, prematurely terminated traversals, and traversals with dead
  code.

  \smallskip

  Based on an inventory of possible programming errors we
  explore options of static typing and static analysis so that some
  categories of errors can be avoided. This exploration generates
  suggestions for improvements to strategy libraries as well as their underlying
  programming languages. Haskell is used for illustrations and
  specifications with sufficient explanations to make the presentation
  comprehensible to the non-specialist. The overall ideas are
  language-agnostic and they are summarized accordingly.

\end{abstract}


\end{frontmatter}


\newpage


\mySection{Introduction}
\label{S:intro}


\introSection{Traversal programming}

In the context of data programming with XML trees, object graphs, and
terms in rewriting or functional programming, consider the general
scenarios of \emph{querying and transforming deeply nested and
  heterogeneously structured data.}  Because of deep nesting and
heterogeneity as well as plain structural complexity, data programming
may benefit from designated idioms and concepts. In this paper, we
focus on the notion of \emph{traversal programming} where
functionality is described in terms of so-called \emph{traversal
  strategies}~\cite{VisserBT98,Essence,BallandBKMR07}: we are
specifically concerned with \emph{programming errors} in this context.


\begin{figure}[t]\small
\begin{boxedminipage}{\hsize}
\begin{boxedminipage}{\hsize}
{\normalsize \emph{\textbf{Illustrative queries}}}
\begin{description}
\item[The abstract syntax of a programming language as a data model] \mbox{}
\begin{enumerate}
\item Determine all recursively defined functions.
\item Determine the nesting depth of a function definition.
\item Collect all the free variables in a given code fragment.
\end{enumerate}
\item[The organizational structure of a company as a data model] \mbox{}
\begin{enumerate}
\item Total the salaries  of all employees.
\item Total the salaries of all employees who are not managers.
\item Calculate the manager / employee ratio in the leaf departments.
\end{enumerate}
\end{description}
\end{boxedminipage}

\begin{boxedminipage}{\hsize}
{\normalsize \emph{\textbf{Illustrative transformations}}}
\begin{description}
\item[The abstract syntax of a programming language as a data model] \mbox{}
\begin{enumerate}
\item Inject logging code around function applications.
\item Perform partial evaluation by constant propagation.
\item Perform unfolding or inlining for a specific function.
\end{enumerate}
\item[The organizational structure of a company as a data model] \mbox{}
\begin{enumerate}
\item Increase the salaries of all employees.
\item Decrease the salaries of all non-top-level managers.
\item Integrate a specific department into the hosting department.
\end{enumerate}
\end{description}
\end{boxedminipage}
\end{boxedminipage}
\caption{Illustrative scenarios for traversal programming.}
\label{F:illu-scenarios}
\figskip
\end{figure}


Let us briefly indicate some application domains for such traversal
programming. To this end, consider the illustrative scenarios for
querying and transforming data as shown in
\autoref{F:illu-scenarios}. The listed queries and transformations
benefit from designated support for traversal programming. The
scenarios assume data models for \emph{a)} the abstract syntax of a
programming language and \emph{b)} the organizational structure of a
company within an information system.  Suitable data models are sketched in
\autoref{F:illu-data} in Haskell syntax.\hsnote{\hs{type} definitions
  are type synonyms. For instance, \hs{Name} is defined to be a
  synonym for Haskell's \hs{String} data type. In contrast, \hs{data}
  types define new algebraic types with one or more constructors (say,
  cases). Consider the data type \hs{Expr}; it declares a number of
  constructors for different expression forms. For instance, the
  constructor \hs{Var} models variables; there is a constructor
  component of type \hs{Name} (in fact, \hs{String}) and hence the
  constructor function's type is \hs{Name -> Expr}.}


\begin{figure}[t]
\begin{boxedminipage}{\hsize}
\begin{boxedminipage}{\hsize}
\textbf{\emph{AST example}: Abstract syntax of a simple functional language}
\begin{lstlisting}[lineskip=3pt]
type Block     = [Function]
data Function  = Function Name [Name] Expr Block
data Expr      = Literal Int
                | Var Name
                | Lambda Name Expr
                | Binary Ops Expr Expr
                | IfThenElse Expr Expr Expr 
                | Apply Name [Expr]
data Ops       = Equal  | Plus | Minus
type Name      = String
\end{lstlisting}
\end{boxedminipage}

\smallskip

\begin{boxedminipage}{\hsize}
\textbf{\emph{Company example}: Organizational structure of a company}
\begin{lstlisting}[lineskip=3pt]
data  Company      = Company [Department]
data  Department   = Department Name Manager [Unit]
data  Manager      = Manager Employee
data  Unit         = EmployeeUnit Employee
                   | DepartmentUnit Department
data  Employee     = Employee Name Salary
type  Name         = String   -- names of employees and departments
type  Salary       = Float    -- salaries of employees
\end{lstlisting}
\end{boxedminipage}
\end{boxedminipage}
\caption{\textbf{Illustrative data models (rendered in
    Haskell). Queries and transformation on such data may require
    traversal programming possibly prone to programming errors.}
  {\small The data models involve multiple types; recursion is
    exercised; there is also a type with multiple alternatives. As a
    result, traversal programmers need to carefully control queries
    and transformations --- leaving room for programming
    errors. \emph{Note on language agnosticism}: Haskell's algebraic
    data types are chosen here without loss of generality. Other
    data-modeling notations such as XML schemas or class diagrams are
    equally applicable.}}
\label{F:illu-data}
\figskip
\end{figure}


A querying traversal essentially visits a compound term, potentially
layer by layer, to extract data from subterms of interest. For
instance, salary terms are extracted from a company term when dealing
with the scenarios of totaling salaries of all or some employees in
\autoref{F:illu-scenarios}.  

A transforming traversal essentially copies a compound term,
potentially layer by layer, except that subterms of interest may be
replaced. For instance, function applications are enriched by logging
code when dealing with the corresponding program transformation
scenario of \autoref{F:illu-scenarios}.


\introSection{Traversal programming with \emph{traversal strategies}}

In this paper, we are concerned with one particular approach to
traversal programming---the approach of \emph{traversal strategies},
which relies on so-called one-layer traversal combinators as well as
other, less original combinators for controlling and composing
recursive traversal. Traversal strategies support an exceptionally
versatile and uniform means of traversal. The term \emph{strategic
  programming} is also in use for this approach to traversal
programming~\cite{Essence}. For example, the first scenario for the
company example requires a traversal strategy as follows:
\begin{itemize}
\item A basic (non-traversing) \emph{rule} is needed to extract salary from
  an employee.
\item The rule is to be iterated over a term by a \emph{traversal scheme}.
\item The traversal scheme itself is defined in terms of \emph{basic combinators}. 
\end{itemize}

The notion of traversal strategies was pioneered by Eelco Visser and
collaborators~\cite{LuttikV97,VisserB98,VisserBT98} in the broader
context of term rewriting. This seminal work also led to Visser et
al.'s Stratego/XT~\cite{Visser03,BravenboerKVV06}---a domain-specific
language, in fact, an infrastructure for software
transformation. Other researchers in the term rewriting and software
transformation communities have also developed related forms of
traversal strategies; see, e.g.,
\cite{BSV97,BorovanskyKKR01,BrandKV03,WinterS04}. 

Historically, strategic programming has been researched considerably
in the context of software transformation.  We refer the reader
to~\cite{VisserBT98,Visser00,Laemmel02-TGR,LaemmelV03-PADL,LiTR05,WinterBFRW06}
for some published accounts of actual applications of traversal
programming with traversal strategies.

The Stratego approach inspired strategic programming approaches for
other programming
paradigms~\cite{LaemmelV02-PADL,Essence,Visser03-PHD,BallandBKMR07,BallandMR08}.
An early functional approach, known by the name `Strafunski',
inspired the `Scrap your boilerplate' (SYB) form of generic
functional
programming~\cite{LaemmelPJ03,LaemmelPJ04,LaemmelPJ05,RenE06,HinzeL06,HinzeLO06},
which is relatively established today in the Haskell community. SYB,
in turn, inspired cross-paradigm variations, e.g., `Scrap your
boilerplate in C++'~\cite{MunkbyPSZ06}.  Traversal strategies \`a la
Stratego were also amalgamated with attribute
grammars~\cite{KatsSV09}.  Another similar form of traversal
strategies are those of the rewriting-based approach of
HATS~\cite{WinterS04,Winter05,WinterB06}. 

There are also several, independently developed forms of traversal
programming: TXL's strategies~\cite{Cordy06}, adaptive traversal
specifications~\cite{LieberherrPO04,LaemmelVV03-AOSD,AbdelmegedL07,AbdelmegedL07},
C$\omega$'s data access~\cite{BiermannMS05}, XPath-like queries and
XSLT transformations~\cite{Laemmel07,CunhaV07} as well as diverse (non-SYB-like)
forms of generic functional
programming~\citeGFP.


\introSection{Research topic: \emph{programming errors} in traversal programming}

Despite the advances in foundations, programming support and
applications of traversal programming with strategies, the use and the
definition of programmable traversal strategies has remained the
domain of the expert, rather than gaining wider usage. We contend that
the principal obstacle to wider adoption is the severity of some
possible pitfalls, which make it difficult to use strategies in
practice. Some of the programming errors that arise are familiar,
e.g., type errors in rewrite rules, but other errors are of a novel
nature. Their appearance can be off-putting to the newcomer to the
field, and indeed limit the productivity  of more experienced
strategists.

Regardless of the specifics of an approach for traversal
programming---the potential for programming errors is relatively
obvious. Here are two possible errors that may arise for the
traversal scenarios of \autoref{F:illu-scenarios}:

\begin{description}

\item[A programming error implying an incorrect result.] In one of the
  scenarios for the company example, a query is assumed to total the
  salaries of all employees who are \emph{not} managers. One approach
  is to use two type-specific cases in the traversal such that
  managers contribute a subtotal of zero whereas regular employees
  contribute with their actual salary. The query should combine the
  type-specific cases in a traversal that ceases when either of the two
  cases is applicable. A traversal could be incorrect in that it
  continues even upon successful application of either case. As a
  result, managers are not effectively excluded from the total because
  the traversal would eventually encounter the employee term inside
  each manager term.

\medskip

\item[A programming error implying divergence.] In one of the scenarios
  for the AST example, a transformation is assumed to \emph{unfold or
    inline a specific function}. This problem involves two traversals:
  one to find the definition of the function, another one to actually
  affect references to the function. Various programming errors are
  conceivable here. For instance, the latter traversal may end up
  unfolding recursive function definitions indefinitely. Instead, the
  transformation should not attempt unfolding for any subterm that was
  created by unfolding itself. This can be achieved by bottom-up
  traversal order.

\end{description}

\introSection{Research objective: \emph{advice on library and language improvements}}

We envisage that the traversal programming of the future will be easier and
safer because it is better understood and better supported. Strategy
libraries and the underlying programming languages need to be improved
in ways that programming errors in the sense of incorrect strategy
application and composition are less likely. Our research objective is
to provide advice on such improvements.

We contend that \emph{static typing} and \emph{static program analysis} are
key tools in avoiding programming errors by detecting favorable or
unfavorable behaviors of strategies statically, that is, without
running the program. Advanced types are helpful in improving libraries
so that their abstractions describe more accurately the contracts for
usage. Static program analysis is helpful in verifying correctness of
strategy composition. 

Accordingly, in this paper, we identify pitfalls of strategic
programming and discover related properties of basic strategy
combinators and common library combinators. Further, we research the
potential of static typing and static program analysis with regard to
our research objective of providing advice on improved libraries and
languages for strategic programming.


\introSection{Summary of the paper's contributions\footnote{\emph{Note
      on the relationship to previous work by the authors}: a short
    version of this paper appeared in the proceedings of the 8th
    Workshop on Language Descriptions, Tools and Applications (LDTA
    2008), published by ENTCS. The present paper is generally more
    detailed and updated, but its distinctive contribution is research
    on static program analysis for traversal strategies in
    \S\ref{S:analysis}; only one of the analyses was sketched briefly
    in the short version. The present paper also takes advantage of
    other previously published work by two of the present authors in
    so far that some of the strategy properties discovered by
    \cite{KaiserL09} help steering research on programming errors in
    the present paper. Programming errors are not systematically
    discussed in \cite{KaiserL09}. Also, type systems and static
    program analysis played no substantial role in said work. Instead,
    the focus was on demonstrating the potential of theorem proving in
    the context of traversal strategies.}}

\begin{enumerate}

\item We provide a fine-grained \emph{inventory of programming errors}
  in traversal programming with traversal strategies; see
  \S\ref{S:errors}. To this end, we reflect on the use of basic
  strategy combinators and library abstractions in the solution
  of traversal problems.

\medskip

\item We explore the \emph{utility of static typing} as means of
  taming traversal strategies; see \S\ref{S:typing}. This exploration
  clarifies what categories of programming errors can be addressed, to
  some extent, by available advanced static type systems. We use
  examples written in Haskell 2010, with extensions, for
  illustrations.

\medskip

\item We explore the \emph{utility of static program analysis} as
  means of taming traversal strategies; see \S\ref{S:analysis}. Static
  analysis is an established technique for dealing with correctness
  and performance properties of software. We study success and failure
  behavior, dead code, and termination.

\end{enumerate}


\introSection{Haskell \emph{en route}}

In this paper, we use Haskell for two major purposes: \emph{a)}
illustrations of static typing techniques for taming strategies;
\emph{b)} specifications of algorithms for static program
analysis. All ideas are also explained informally and specific Haskell
idioms are explained---as they are encountered---so that the
presentation should also be comprehensible to the non-specialist.  The
overall ideas of taming strategies by static typing and static program
analysis are not tied to Haskell; they are largely agnostic to the
language---as long as it supports strategic programming style.
Accordingly, all subsections of \S\ref{S:typing} and
\S\ref{S:analysis} begin with a language-agnostic, informal advice for
improving strategy libraries and the underlying languages.


\introSection{Scope of this work}

Our work ties into traversal strategies \`a la Stratego (also \`a la
Strafunski and `Scrap Your Boilerplate') in that it intimately
connects to idiomatic and conceptual details of this class of
approaches: the use of one-layer traversal combinators, the style of
mixing generic and problem-specific behavior, the style of controlling
traversal through success and failure, and a preference for certain
important traversal schemes.

Nevertheless, our research approach and some of the results may be
applicable to other forms of traversal programming, e.g., adaptive
programming~\citeAP, (classic) generic functional
programming~\citeGFP, XML queries or transformations (based on
mainstream languages). However, we do not explore such potential. As a
brief indication of potential relevance, let us consider
XSLT~\cite{XSLT} with its declarative processing model for XML that
can be used to implement transformations between XML-based
documents. An XSLT transformation consists of a collection of template
rules, each of which describes which XML (sub-)trees to match and how to
process them. Processing begins at the root node and proceeds by
applying the best fitting template pattern, and recursively processing
according to the template. Hence, traversal is implicit in the XSLT
processing model, but the templates control the traversal. Several
problems that we study in the present paper---such as dead code
or divergence---can occur in XSLT and require similar analyses.


\introSection{Road map of the paper}

\begin{itemize}
\item \S\ref{S:background} provides background on strategic programming.
\item \S\ref{S:errors} makes an inventory of programming errors
  in strategic programming.
\item \S\ref{S:typing} researches the potential of static typing for
  taming traversal strategies.
\item \S\ref{S:analysis} researches the potential of static analysis for
  taming traversal strategies.
\item \S\ref{S:related} discusses related work.
\item \S\ref{S:concl} concludes the paper.
\end{itemize}

\mySection{Background on strategic programming}
\label{S:background}

This section introduces the notion of traversal strategy and the style
of strategic programming. We integrate basic material that is
otherwise scattered over several publications. The collection is also
valuable in so far that we make an effort to point out properties of
strategies as they will be useful eventually in the discussion
of programming errors.

We describe traversal strategies in three different styles.  First, we
use a formal style based on a grammar and a natural semantics.  Second,
we use an interpreter style such that the semantics is essentially
implemented in Haskell. Third, we embed strategies into Haskell---this
is the style that is useful for actual strategic programming (in
Haskell) as opposed to formal investigation.


\begin{figure}[t!]
\begin{boxedminipage}{\hsize}
{\footnotesize
\[\begin{array}{lcll}
s & ::= & t \rewritesto t &
\smbox{(Rewrite rules as basic building blocks.)}
\\
  & |   & \idT  &
\smbox{(Identity transformation; succeeds and returns input.)}
\\
  & |   & \failT  &
\smbox{(Failure transformation; fails and returns failure ``$\failure$''.)}
\\
  & |   & \sequ{s}{s} \ \ \ \ \ \ \ \ \ \ \  &
\smbox{(Left-to-right sequential composition.)}
\\
  & |   & \choice{s}{s} &
\smbox{(Left-biased choice; try left argument first.)}
\\
  & |   & \all{s} &
\smbox{(Transform all immediate subterms by $s$; maintain constructor.)}
\\
  & |   & \one{s} &
\smbox{(Transform one immediate subterm by $s$; maintain constructor.)}
\\
  & |   & v & 
\smbox{(A variable strategy subject to binding or substitution.)}
\\
  & |   & \rec{v}{s} &
\smbox{(Recursive closure of $s$ referring to itself as $v$.)}
\end{array}\]
}
\end{boxedminipage}
\caption{\textbf{Syntax of strategy primitives for transformations.}}
\label{F:syntax}
\figskip
\end{figure}


\mySubsection{A core calculus of transformations}
\label{S:core}

\autoref{F:syntax} shows the syntax of a core calculus for
transformations. Later we will also cover queries.\footnote{\emph{Note
    on terminology}: some of the strategic programming literature
  prefers the terms `type-preserving strategies' for transformations
  and `type-unifying strategies' for queries. We do not follow this
  tradition here---in the interest of a simple terminology.}
Transformations are sufficient to introduce strategies in a formal
manner; we are confident that our approach easily extends to
queries. The core calculus of \autoref{F:syntax} follows closely
Visser \emph{et al.}'s seminal work~\cite{VisserBT98}.

The calculus contains basic strategies as well as strategy
\emph{combinators}.  The basic strategy \idT{} denotes the always
succeeding identity transformation, which can be thought of as the
identity function. The basic strategy \failT{} denotes the always
failing transformation. Here we note that strategies can either fail,
succeed, or diverge. (We use the symbol ``\failure'' to denote failure
in the upcoming semantics.) There is also a basic strategy form for
rewrite rules in the sense of term rewriting. All
strategies---including rewrite rules---are applied at the root node of
the input. Traversal into internal nodes of the tree is provided by
designated combinators.

Turning to the combinators, the composed strategy $\sequ{s}{s'}$
denotes the sequential composition of $s$ and $s'$. The composed
strategy $\choice{s}{s'}$ denotes left-biased choice: try $s$ first,
and try $s'$ second---if $s$ failed.  The characteristic
combinators of strategic programming are \allchar{} and \onechar{}
(also known as `\hs{all}' and `\hs{one}'). These combinators model
so-called \emph{one-layer traversal}. The strategy \all{s} applies $s$
to all immediate subterms of a given term. If there is any subterm for
which $s$ fails, then \all{s} fails, too. Otherwise, the result is the
term that is obtained by the application of the original outermost
constructor to the processed subterms. The strategy \one{s} applies
$s$ only to one immediate subterm, namely the leftmost one, if any,
for which $s$ succeeds. If $s$ fails for all subterms, then \one{s}
fails, too. The one-layer traversal combinators, when used within a
recursive closure, enable the key capability of strategic programming:
to describe arbitrarily deep traversal into heterogeneously typed
terms.

A comprehensive instantiation of strategic programming may require
additional strategy primitives that we omit here. For instance,
Stratego also provides strategy forms for congruences (i.e., the
application of strategies to the immediate subterms of specific
constructors), tests (i.e., a strategy is tested for success, but its
result is not propagated), and negation (i.e., a strategy to invert
the success and failure behavior of a given
strategy)~\cite{VisserBT98}. As motivated earlier, we also omit
specific primitives for queries.


\begin{figure}[t!]
\begin{boxedminipage}{\hsize}

{\small

\ir{rule^+}{%
\exists \theta.\ (\theta(t_l) = t \wedge \theta(t_r) = t')
}{%
\ioj{\apply{t_l \rewritesto t_r}{t}}{t'}
}

\ax{id^+}{%
\ioj{\apply{\idT}{t}}{t}
}

\ir{sequ^+}{%
\ioj{\apply{s_1}{t}}{t'}
\dnp
\ioj{\apply{s_2}{t'}}{t''}
}{%
\ioj{\apply{\sequ{s_1}{s_2}}{t}}{t''}
}

\ir{choice^+.1}{%
\ioj{\apply{s_1}{t}}{t'}
}{%
\ioj{\apply{\choice{s_1}{s_2}}{t}}{t'}
}

\ir{choice^+.2}{%
\ioj{\apply{s_1}{t}}{\failure}
\dnp
\ioj{\apply{s_2}{t}}{t'}
}{%
\ioj{\apply{\choice{s_1}{s_2}}{t}}{t'}
}

\ir{all^+}{%
\forall i \in \{1,\ldots,n\}.\ \ioj{\apply{s}{t_i}}{t'_i}
}{%
\ioj{\apply{\all{s}}{c(t_1, \ldots, t_n)}}{c(t'_1, \ldots, t'_n)}
}

\ir{one^+}{%
\begin{array}{l}
\exists i \in \{1,\ldots,n\}.\\
\ \begin{array}{ll}
& \ioj{\apply{s}{t_i}}{t'_i}\\
\wedge & \forall i' \in \{1,\ldots,i-1\}.\ \ioj{\apply{s}{t_{i'}}}{\failure}\\
\wedge & \forall i' \in \{1,\ldots,i-1,i+1,\ldots,n\}.\ t_{i'} = t'_{i'}
\end{array}
\end{array}
}{%
\ioj{\apply{\one{s}}{c(t_1, \ldots, t_n)}}{c(t'_1, \ldots, t'_n)}
}

\ir{rec^{+}}{%
\ioj{\apply{s[v\mapsto\rec{v}{s}]}{t}}{t'}
}{%
\ioj{\apply{\rec{v}{s}}{t}}{t'}
}

}

\end{boxedminipage}

\caption{\textbf{Positive rules of natural semantics for the core calculus.}}
\label{F:plus}
\figskip
\end{figure}


\begin{figure}[t!]
\begin{boxedminipage}{\hsize}

{\small

\ir{rule^-}{%
\not\exists \theta.\ \theta(t_l) = t
}{%
\ioj{\apply{t_l \rewritesto t_r}{t}}{\failure}
}

\ax{fail^-}{%
\ioj{\apply{\failT}{t}}{\failure}
}

\ir{seq^-.1}{%
\ioj{\apply{s_1}{t}}{\failure}
}{%
\ioj{\apply{\sequ{s_1}{s_2}}{t}}{\failure}
}

\ir{seq^-.2}{%
\ioj{\apply{s_1}{t}}{t'}
\dnp
\ioj{\apply{s_2}{t'}}{\failure}
}{%
\ioj{\apply{\sequ{s_1}{s_2}}{t}}{\failure}
}

\ir{choice^-}{%
\ioj{\apply{s_1}{t}}{\failure}
\dnp
\ioj{\apply{s_2}{t}}{\failure}
}{%
\ioj{\apply{\choice{s_1}{s_2}}{t}}{\failure}
}

\ir{all^-}{%
\exists i \in \{1,\ldots,n\}.\ \ioj{\apply{s}{t_i}}{\failure}
}{%
\ioj{\apply{\all{s}}{c(t_1, \ldots, t_n)}}{\failure}
}

\ir{one^-}{%
\forall i \in \{1,\ldots,n\}.\ \ioj{\apply{s}{t_i}}{\failure}
}{%
\ioj{\apply{\one{s}}{c(t_1, \ldots, t_n)}}{\failure}
}

\ir{rec^{-}}{%
\ioj{\apply{s[v\mapsto\rec{v}{s}]}{t}}{\failure}
}{%
\ioj{\apply{\rec{v}{s}}{t}}{\failure}
}

}

\end{boxedminipage}

\caption{\textbf{Negative rules of natural semantics for the core calculus.}}
\label{F:minus}
\figskip
\end{figure}


Following \cite{VisserBT98} and subsequent work on the formalization
of traversal strategies~\cite{Laemmel03-JLAP,KaiserL09}, we give the
formal semantics of the core calculus as a big-step
operational semantics using a success and failure model, shown in
\autoref{F:plus} and \autoref{F:minus}.  (The version shown has been
extracted from a mechanized model of traversal strategies, based on
the theorem prover Isabelle/HOL~\cite{KaiserL09}.) 

The judgement \ioj{\apply{s}{t}}{r} describes a relation between a
strategy expression $s$, an input term $t$, and a result $r$, which is
either a proper term or failure---denoted by `\failure'.  Based on
tradition, we separate positive rules (resulting in a proper term) and
negative rules (resulting in `\failure'). Incidentally, this
distinction already helps in understanding the success and failure
behavior of strategies. For instance, the positive rule \tg{all^+}
models that the strategy application \apply{\all{s}}{c(t_1, \ldots,
  t_n)} applies $s$ to all the $t_i$ such that new terms $t'_i$ are
obtained and used in the result term $c(t'_1, \ldots, t'_n)$. The
negative rule covers the case that at least one of the applications
\apply{s}{t'_i} resulted in failure.

The semantics shown uses variables only for the sake of recursive
closures, and the semantics of these is modelled by
substitution. We could also furnish variables for the sake of
parametrized strategy expressions, i.e., binding blocks of strategy
definitions, as this is relevant for reusability in practice, but we
do not furnish this elaboration of the semantics here for brevity's
sake.

The semantics commits to a simple, deterministic model: each strategy
application evaluates to one term deterministically, or it fails, or
it diverges. Further, the semantics of the choice combinator is
left-biased with only local backtracking. Non-deterministic semantics
have been considered elsewhere such that results may be lists (or sets)
of terms with the empty list (or set) representing failure
\cite{BorovanskyKKMR98,BorovanskyKKR01}. Non-determinism is also
enabled naturally by a monadic-style functional programming embedding
when using the list monad.


\begin{figure}[t!]
\begin{boxedminipage}{\hsize}
\vspace{-55\in}
{\small\[\begin{array}{lcll}
  \w{full\_td}(s) & = & \rec{v}{\sequ{s}{\all{v}}} & \mbox{\scriptsize-\,-
    Apply $s$ to each subterm in top-down manner}\\
  \w{full\_bu}(s) & = & \rec{v}{\sequ{\all{v}}{s}} & \mbox{\scriptsize-\,-
    Apply $s$ to each subterm in bottom-up manner}\\
  \w{once\_td}(s) & = & \rec{v}{\choice{s}{\one{v}}} & \mbox{\scriptsize-\,-
    Find one subterm (top-down) for which $s$ succeeds}\\
  \w{once\_bu}(s) & = & \rec{v}{\choice{\one{v}}{s}} & \mbox{\scriptsize-\,-
    Find one subterm (bottom-up) for which $s$ succeeds}\\
  \w{stop\_td}(s) & = & \rec{v}{\choice{s}{\all{v}}} & \mbox{\scriptsize-\,-
    Stops when $s$ succeeds on a `cut' through the tree}\\
  \w{stop\_bu}(s) & = & \rec{v}{\choice{\all{v}}{s}} & \mbox{\scriptsize-\,-
    An illustrative programming error; see the text.}\\
  \w{innermost}(s) & = & \w{repeat}(\w{once\_bu}(s)) &
  \mbox{\scriptsize-\,-
    A form of innermost normalization \`a la term rewriting.}\\
  \w{repeat}(s) & = & \rec{v}{\w{try}(\sequ{s}{v})} &
  \mbox{\scriptsize-\,-
    Fixed point iteration; apply $s$ until it fails.}\\
  \w{try}(s) & = & \choice{s}{\idT} &
  \mbox{\scriptsize-\,-
    Recovery from failure of $s$ with catch-all \idT.}
\end{array}\]
}
\vspace{-55\in}
\end{boxedminipage}
\caption{\textbf{Familiar traversal schemes.}}
\label{F:schemes}
\figskip
\end{figure}


\mySubsection{Traversal schemes}
\label{S:schemes}

Assuming an \emph{ad hoc} notation for parametrized strategy
definition---think, for example, of macro expansion---some familiar
traversal schemes and necessary helpers are defined in
\autoref{F:schemes}.  We refer to \cite{Laemmel02-WRS,RenE06} for a
more detailed discussion of the design space for traversal schemes.

The traversal scheme $\w{stop\_bu}(s)$ from \autoref{F:schemes} is in
fact bogus; it is only included here to provide an early, concrete
example of a conceivable programming error. That is, the argument $s$
will never be applied. Instead, any application of the scheme will
simply perform a deep identity traversal. This property can be proven
with relatively little effort by induction on the structure of terms,
also using the auxiliary property that \all{s} is the identity for all
constant terms, i.e., terms without any subterms, no matter what $s$
is. A library author may notice the problem eventually. A `regular'
strategic programmer may make similar programming errors when
developing problem-specific traversals.


\begin{figure}[t!]
\begin{boxedminipage}{\hsize}
\vspace{-55\in}
{\small
\[\begin{array}{lllclcl}
\mbox{[unit of ``;'']} &\ \ \ \ & \sequ{\idT}{s} & = & s & = & \sequ{s}{\idT}\\
\mbox{[zero of ``;'']} & & \sequ{\failT}{s} & = & \failT & = & \sequ{s}{\failT}\\
\mbox{[unit of ``\choicechar'']} & & \choice{\failT}{s} & = & s & = & \choice{s}{\failT}\\
\mbox{[left zero of ``\choicechar'']} & & \choice{\idT}{s} & = & \idT\\
\mbox{[associativity of ``;'']} & & \sequ{s_1}{(\sequ{s_2}{s_3})} & = & \sequ{(\sequ{s_1}{s_2})}{s_3}\\
\mbox{[associativity of ``\choicechar'']} & & \choice{s_1}{(\choice{s_2}{s_3})} & = & \choice{(\choice{s_1}{s_2})}{s_3}\\
\mbox{[left distributivity]} & & \sequ{s_1}{(\choice{s_2}{s_3})} & = & \choice{(\sequ{s_1}{s_2})}{(\sequ{s_1}{s_3})}\\
\mbox{[one-layer identity]} & & \all{\idT} & = & \idT\\
\mbox{[one-layer failure]} & & \one{\failT} & = & \failT\\
\mbox{[fusion law]} & & \sequ{\all{s_1}}{\all{s_2}} & = & \all{\sequ{s_1}{s_2}}\\
\mbox{[``\allchar'' with a constant]} & & \w{constant}(t) &
\Rightarrow & \apply{\all{s}}{t} & = & t\\
\mbox{[``\onechar'' with a constant]} & & \w{constant}(t) &
\Rightarrow & \apply{\one{s}}{t} & = & \failure\\
\mbox{[``\allchar'' with a non-constant]} & & \neg\w{constant}(t) &
\Rightarrow & \apply{\all{\failT}}{t} & = & \failure\\
\mbox{[``\onechar'' with a non-constant]} & & \neg\w{constant}(t) &
\Rightarrow & \apply{\one{\idT}}{t} & = & t \\ \hline\hline
\mbox{[\sout{commutativity of ``;''}]} & & \sequ{s}{s'} & \neq & \sequ{s'}{s}\\
\mbox{[\sout{commutativity of ``\choicechar''}]} & & \choice{s}{s'} & \neq & \choice{s'}{s}\\
\mbox{[\sout{right distributivity}]} & & \sequ{(\choice{s_1}{s_2})}{s_3} & \neq & \choice{(\sequ{s_1}{s_3})}{(\sequ{s_2}{s_3})}
\end{array}\]}
\vspace{-55\in}
\end{boxedminipage}

\caption{\textbf{Algebraic laws and non-laws of strategy primitives.}
  In a few laws, in fact, implications, we use an auxiliary judgement
  \w{constant} that holds for all constant terms, i.e., terms with 0
  subterms.}
\label{F:laws}
\figskip
\end{figure}


\mySubsection{Laws and properties}
\label{S:properties}

\autoref{F:laws} lists algebraic laws obeyed by the strategy
primitives. These laws should be helpful in understanding the
primitives and their use in traversal schemes. We refer to
\cite{KaiserL09} for a mechanized model of traversal strategies, which
proves these laws and additional properties. These laws provide
intuitions with regard to the success and failure behavior of
strategies, and they also hint at potential sources of dead code.

For instance, the fusion law states that two subsequent `\allchar'
traversals can be composed into one. Such a simple law does not hold
for `\onechar', neither does it hold generally for traversal
schemes. In fact, little is known about algebraic laws for traversal
schemes, but see~\cite{JohannV03,Reig04} for some related research.

We also illustrate three non-laws at the foot of \autoref{F:laws},
which show putative equalities. The first reflects that fact that
sequential composition is (of course) not commutative, the second that
choice is (indeed) left-biased, and the third that distributivity is
limited. Finding a counterexample to the third non-law we leave as an
exercise for the reader.

Let us also discuss basic properties of success and failure behavior
for strategies. According to~\cite{KaiserL09}, we say that a strategy
$s$ is \emph{infallible} if it does not possibly fail, i.e., for any
given term, it either succeeds or diverges; otherwise $s$ is
fallible. The following properties hold~\cite{KaiserL09}:
\begin{itemize}
\item If $s$ is infallible, then \hs{full_td}\ $s$ and \hs{full_bu}\ $s$ are infallible.
\item No matter the argument $s$, \hs{stop_td}\ $s$ and
  \hs{innermost}\ $s$ are infallible.
\item No matter the argument $s$, \hs{once_td}\ $s$ and
  \hs{once_bu}\ $s$ are fallible.
\end{itemize}

These properties can be confirmed based on induction arguments. For
instance, (the all-based branch of the choice in) a stop-top-down
traversal succeeds eventually for `leaves', i.e., terms without
subterms, and it succeeds as well for every term for which the
traversal succeeds for all immediate subterms. Hence, by induction
over the depth of the term, the traversal succeeds universally.

The discussion of termination behavior for traversal schemes is more
complicated, but let us provide a few intuitions here. That is, it is
easy to see that full bottom-up traversal converges---as long as its
argument strategy does not diverge---because the scheme essentially
performs structural recursion on the input term. In contrast, full
top-down traversal may diverge rather easily because the argument
strategy could continuously increase the given term before traversing
into it. This will be illustrated in \S\ref{S:errors}. We will address
termination by means of static program analysis in \S\ref{S:analysis}.


\begin{figure}[t!]
\begin{boxedminipage}{\hsize}
\begin{lstlisting}[aboveskip=3pt]
data T x
 = Id
 | Fail
 | Seq (T x) (T x)
 | Choice (T x) (T x)
 | Var x
 | Rec (x -> T x)
 | All (T x)
 | One (T x)
\end{lstlisting}
\end{boxedminipage}
\caption{\textbf{Syntax according to \autoref{F:syntax} in Haskell.}}
\label{F:syntax-hs}
\figskip
\end{figure}


\mySubsection{A Haskell-based interpreter for strategies}
\label{S:interpreter}

Let us also provide an interpreter-based model of the core calculus in
Haskell. In fact, we provide this model in a way that we can easily
refine it later on for the purpose of static program analysis in
\S\ref{S:analysis}. The interpreter-based model is not very useful
though for actual programming in Haskell because it is essentially
untyped with regard to the terms being transformed. We will properly
embed strategies in Haskell in \S\ref{S:embedding}.

\autoref{F:syntax-hs} shows the algebraic data type \hs{T} for the
syntax of transformations according to the core calculus. There are
constructors for the various basic strategies and combinators. There
is, however, no strategy form for rewrite rules because we can easily
represent rewrite rules as functions. Please note that type \hs{T} is
parameterized by the type \hs{x} for variables. Here we apply a
modeling technique for syntax that allows us to model variables and
binding constructs of the interpreted language with variables and
binding constructs of the host language. Further, this particular
style supports repurposing this syntax most conveniently during
abstract interpretation in \S\ref{S:analysis}. Finally, the style
frees us from implementing substitution.


\begin{figure}[t!]
\begin{boxedminipage}{\hsize}
\begin{lstlisting}[aboveskip=3pt]
-- Representation of terms
data Term   = Term Constr [Term]
type Constr = String

-- The semantic domain for strategies
type Meaning = Term -> Maybe Term

-- Interpreter function
interpret :: T Meaning -> Meaning
interpret Id            = Just
interpret Fail          = const Nothing
interpret (Seq s s')    = maybe Nothing (interpret s') . interpret s
interpret (Choice s s') = \t -> maybe (interpret s' t) Just (interpret s t)
interpret (Var x)       = x
interpret (Rec f)       = fixProperty (interpret . f)
interpret (All s)       = transform (all (interpret s))
interpret (One s)       = transform (one (interpret s))

-- Fixed-point combinator
fixProperty :: (x -> x) -> x
fixProperty f = f (fixProperty f)

-- Common helper for All and One
transform :: ([Term] -> Maybe [Term]) -> Meaning
transform f (Term c ts)
 = maybe Nothing (Just . Term c) (f ts)

-- Transform all terms in a list
all :: Meaning -> [Term] -> Maybe [Term]
all f ts = kids ts'
 where
  ts' = map f ts
  kids [] = Just []
  kids (Just t':ts') = maybe Nothing (Just . (:) t') (kids ts')

-- Transform one term in a list
one :: Meaning -> [Term] -> Maybe [Term]
one f ts =  kids ts ts'
 where
  ts' = map f ts
  kids [] [] = Nothing
  kids (t:ts) (Nothing:ts') = maybe Nothing (Just . (:) t) (kids ts ts')
  kids (_:ts) (Just t':ts') = Just (t':ts)
\end{lstlisting}
\end{boxedminipage}
\caption{\textbf{Haskell-based interpreter for transformation strategies.}}
\label{F:interpreter}
\end{figure}


\autoref{F:interpreter} shows the actual interpreter, which is
essentially a recursive function, \hs{interpret}, on the syntactical
domain---subject to a few auxiliary declarations as follows. There is
an algebraic data type \hs{Term} for terms to be transformed;
constructors are assumed to be strings; see the type synonym
\hs{Constr}. There is a type synonym \hs{Meaning} to make explicit the
type of functions computed by the interpreter. That is, a strategy is
mapped to a function from terms to terms where the function is partial
in the sense of the \hs{Maybe} type constructor.\hsnote{\hs{Maybe} is
  defined as follows: \hs{data Maybe x = Just x | Nothing}. The
  constructor \hs{Just} is applied when a value (i.e., a result in our
  case) is available whereas the constructor \hs{Nothing} is applied
  otherwise. Maybe values can be inspected by regular pattern
  matching, but we also use the convenience function \hs{maybe :: b ->
    (a -> b) -> Maybe a -> b} which applies the first argument if the
  given `maybe' is \hs{Nothing} and otherwise the second argument (a
  function) to the value.} The type \hs{Meaning} hence also describes
what variables are to be bound to.\hsnote{Haskell in all its glory has
  infinite and partial data structures, such as trees with undefined
  leaves, or indeed undefined subtrees. In principle, the data type
  \hs{Term} can be used in such a manner. In the presence of infinite
  and partial structures, the discussion of strategy semantics and
  properties (most notably, termination) becomes more subtle. In this
  paper, we are limiting our discussion to finite, fully defined
  data. (The subject of coinductive strategies over coinductive types
  may be an interesting topic for future work.)  We also skip over the
  issues of laziness in most cases.}

The equations of the \hs{interpret} function combine the positive and
negative rules of the natural semantics in a straightforward
manner. The only special case is the approach to recursion. We use a
fixed point combinator, \hs{fixProperty}, to this end. Its name
emphasizes that the definition of the operator is immediately the
defining property of a fixed point.

The traversal combinators are interpreted with the help of auxiliary
functions \hs{transform}, \hs{all}, and \hs{one}. One-layer traversal
is essentially modeled by means of mapping over the list of subterms
while using the folklore list-processing function \hs{map}. The
auxiliary functions are otherwise only concerned with the manipulation
of failure for processed subterms.


\begin{figure}[t!]
\begin{boxedminipage}{\hsize}
\begin{lstlisting}[aboveskip=3pt]
full_bu s   = Rec (\x -> Seq (All (Var x)) s)
full_td s   = Rec (\x -> Seq s (All (Var x)))
once_bu s   = Rec (\x -> Choice (One (Var x)) s)
once_td s   = Rec (\x -> Choice s (One (Var x)))
stop_td s   = Rec (\x -> Choice s (All (Var x)))
innermost s = repeat (once_bu s)
try s       = Choice s Id
repeat s    = Rec (\x -> try (Seq s (Var x)))
\end{lstlisting}
\end{boxedminipage}
\caption{\textbf{Familiar traversal schemes for interpretation in Haskell.}}
\label{F:schemes-hs}
\figskip
\end{figure}


\autoref{F:schemes-hs} shows familiar traversal schemes in the
Haskell-based syntax.


\begin{figure}[t!]
\begin{boxedminipage}{\hsize}
\begin{lstlisting}[columns=fullflexible,aboveskip=3pt]
-- Transformations as generic functions
type T m = forall x. Term x => x -> m x

-- Combinator types
idT :: Monad m => T m
failT :: MonadPlus m  => T m
sequT :: Monad m => T m -> T m -> T m
choiceT :: MonadPlus m  => T m -> T m -> T m
allT :: Monad m => T m -> T m
oneT :: MonadPlus m => T m -> T m
adhocT :: (Term x, Monad m) => T m -> (x -> m x) -> T m

-- Trivial combinator definitions
idT = return
failT = const mzero
sequT f g x = f x >>= g
choiceT f g x = f x `mplus` g x

-- Non-trivial combinator definitions using pseudo-code
allT f (C t1 ... tn) =
    f t1 >>= \t1' -> ... f tn >>= \tn' ->
    return (C t1' ... tn') 

oneT f (C t1 ... tn) =
    ... -- elided for brevity

adhocT s f x =
    if typeOf x == argumentTypeOf f 
       then f x
       else s x
\end{lstlisting}
\end{boxedminipage}
\caption{\textbf{Embedding strategies (transformations) in Haskell.}}
\label{F:T}
\figskip
\end{figure}


\begin{figure}[t!]
\begin{boxedminipage}{\hsize}
\small
A Haskell class is specified by a signature, which can be thought of as an interface. An implementation of the class---called an \emph{instance} in Haskell---is given by defining the interface functions for a particular type. A typical example is the equality class, \hs{Eq}, shown with an instance for the Boolean type, \hs{Bool}. 
\begin{lstlisting}
class Eq a where
  (==) :: a -> a -> Bool
instance Eq Bool where
  (x == y)  =  if x then y else (not y)
\end{lstlisting}
A function may be polymorphic yet require that a type is a member of a particular class, as in the list membership function;  the context \hs{Eq a => ...} constrains \hs{a} to be a member of the \hs{Eq} class.
\begin{lstlisting}
elem :: Eq a => a -> [a] -> Bool
elem x ys = or [ x==y | y<-ys ]
\end{lstlisting}
Since \hs{Bool} is in the class \hs{Eq}, then \hs{elem} can be used over Boolean lists.
\end{boxedminipage}

\caption{\textbf{A note on classes in Haskell.}}
\label{F:classes}
\figskip
\end{figure}


\begin{figure}[t!]
\begin{boxedminipage}{\hsize}
  \small Classes can also describe interfaces over type constructors,
  that is, functions from types to types. For instance: 
\begin{lstlisting}
class Monad m where
  return :: a -> m a
  (>>=)  :: m a -> (a -> m b) -> m b

class Monad m => MonadPlus m where
  mzero :: m a
  mplus  :: m a -> m a -> m a
\end{lstlisting}
Monads encapsulate an interface for `computations over \hs{a}', since
\hs{return x} gives the trivial computation of the value \hs{x} and
\hs{(>>=)} or `bind' allows computations to be sequenced. The simplest
implementation of \hs{Monad} is the identity type function:
\begin{lstlisting}
data Id a = Id { getId :: x }

instance Monad Id where
  return x = Id x
  c >>= f = f (getId c)
\end{lstlisting}

\smallskip
\noindent
Other instances of \hs{Monad} provide for non-deterministic or
stateful computation, which can be used to good effect in traversals,
e.g., to accumulate context information during the traversal.

\smallskip
\noindent
In a similar way, \hs{MonadPlus} encapsulates the concept of
computations that might fail, witnessed by the \hs{mzero} binding, and
\hs{mplus} combines together the results of two computations that
might fail, transmitting failure as appropriate. The simplest instance
of \hs{MonadPlus} is the \hs{Maybe} type:
\begin{lstlisting}
data Maybe a = Nothing | Just a

instance Monad Maybe where
  return x = Just x
  Nothing >>= f = Nothing
  (Just x) >>= f = f x

instance MonadPlus Maybe where
  mzero = Nothing
  mplus Nothing y = y
  mplus x _ = x
\end{lstlisting}
Every instance of \hs{MonadPlus} presupposes an instance of \hs{Monad},
but not \emph{vice versa}.
\end{boxedminipage}
\caption{\textbf{A note on monads in Haskell.}}
\label{F:monads}
\figskip
\end{figure}


\mySubsection{Embedding strategies into Haskell}
\label{S:embedding}

By embedding strategies into Haskell, they can be applied to
programmer-defined data types such as those illustrative data models
for companies and programming-language syntax in \autoref{F:illu-data}
in the introduction.

\autoref{F:T} defines the generic function type for transformations
and the function combinators for the core calculus. A number of
aspects of this embedding need to be carefully motivated.

The type \hs{T} uses forall-quantification to emphasize that
strategies are indeed generic (say, polymorphic) functions because
they are potentially applied to many different types of subterms along
traversal. The context \hs{Term x => ...} emphasizes that strategies
are not universally polymorphic functions, but they can only be
applied to types that instantiate the Haskell class
\hs{Term}.\hsnote{\autoref{F:classes} gives a brief overview of Haskell
  classes (say, type classes) for those unfamiliar with this aspect of
  the Haskell language.} Thus, \hs{Term} is the \emph{set} of types of
terms. (For comparison, in the interpreter-based model, \hs{Term}
denotes the type of terms.) The operations of the \hs{Term} class
enable traversal and strategy extension (see below). The specifics of
these operations are not important for the topic of the present paper;
see the Strafunski/`Scrap Your Boilerplate'
literature~\cite{LaemmelV02-PADL,LaemmelPJ03} for details.

There is the \hs{adhocT} combinator that has no counterpart in the
formal semantics and the interpreter-based model because it is
specifically needed for static typing when different types can be
traversed. The \hs{adhocT} combinator enables so-called \emph{strategy
  extension} as follows. The strategy \hs{adhocT}\ $s$\ $f$ constructs
a new strategy from the given strategy $s$ such that the result
behaves like the type-specific case $f$, when $f$ is applicable and
like $s$ otherwise. This is also expressed by the pseudo-code. We omit
technical details that are not important for the topic of the present
paper; see, again, \cite{LaemmelV02-PADL,LaemmelPJ03} for details. It
is important though to understand that an operation like \hs{adhocT}
combinator is essential for a (conservative) typeful embedding. This
will be illustrated shortly.

When compared to the interpreter of \autoref{F:interpreter}, the
embedding does not refer to the \hs{Maybe} type constructor for the
potential of failure in strategy application. Instead, a
type-constructor parameter for a monad \hs{m} is
used.\hsnote{\autoref{F:monads} gives a brief overview of monads for
  those unfamiliar with the concept.} The use of monads is a strict
generalization of the use of \hs{Maybe} because \emph{a)} \hs{Maybe}
is a specific monad and \emph{b)} other monads can be chosen to
compose traversal behavior with additional computational aspects. This
generalization has been found to be essential in practical, strategic
programming in Haskell. By instantiating \hs{m} as \hs{Maybe}, we get
these types:

\begin{lstlisting}[columns=fullflexible]
idT :: T Maybe
failT :: T Maybe
sequT :: T Maybe -> T Maybe -> T Maybe
...
\end{lstlisting}
We note that the choice of \hs{Monad} versus \hs{MonadPlus} in the
original function signatures of \autoref{F:T} simply expresses what is
required by the combinators' definitions.  For instance, \hs{idT} does
not refer to any members of the \hs{MonadPlus} class whereas
\hs{choiceT} does.

The pseudo-code for the \hs{allT} combinator expresses that the
argument function (say, strategy) is applied to all immediate
subterms, the various computations are sequenced, and a term with the
original outermost constructor is constructed from the intermediate
results---unless failure occurred.


\begin{figure}[t!]
\begin{boxedminipage}{\hsize}
\begin{lstlisting}[aboveskip=7pt]
full_td, full_bu       :: Monad m       => T m -> T m
once_td, once_bu       :: MonadPlus m   => T m -> T m
stop_td                :: MonadPlus m   => T m -> T m
innermost, repeat, try :: MonadPlus m   => T m -> T m

full_td s    = s `sequT` allT (full_td s)
full_bu s    = allT (full_bu s) `sequT` s
once_td s    = s `choiceT` oneT (once_td s)
once_bu s    = oneT (once_bu s) `choiceT` s
stop_td s    = s `choiceT` allT (stop_td s)
innermost s  = repeat (once_bu s)
repeat s     = try (s `sequT` repeat s)
try s        = s `choiceT` idT
\end{lstlisting}
\end{boxedminipage}
\caption{\textbf{Familiar traversal schemes embedded in Haskell.}}
\label{F:schemes-sf}
\figskip
\end{figure}


\autoref{F:schemes-sf} expresses familiar traversal schemes with the
embedding. The function definitions are entirely straightforward, but
two details are worth noticing as they relate to the central topic of
programming errors. First, the function definitions use general
recursion, thereby implying the potential for divergence. Second, the
distinction of \hs{Monad} and \hs{MonadPlus} in the function
signatures signals whether control flow can be affected by fallible
arguments of the schemes, but the types do not imply rejection of
infallible arguments, thereby implying the potential for degenerated
control flow.


\begin{figure}[t!]
\begin{boxedminipage}{\hsize}
\begin{lstlisting}[columns=fullflexible,aboveskip=7pt]
-- Increase the salaries of all employees.
increase_all_salaries :: T Id
increase_all_salaries = full_td (adhocT idT f)
 where
  f (Employee n s) = Id (Employee n (s+1))
\end{lstlisting}
\end{boxedminipage}
\caption{\textbf{Implementation of a transformation scenario from \autoref{F:illu-scenarios}.}}
\label{F:increase}
\figskip
\end{figure}


\autoref{F:increase} composes a traversal for the transformation of
companies such that the salaries of all employees are increased by 1
Euro. The implementation of the traversal is straightforward: we
select a scheme for \emph{full} traversal such that we reach each
node, and we extend the polymorphic identity function with a
monomorphic function for employees so that we increase (in fact,
increment) their salary components. The local function \hs{f} can be
viewed as a rewrite rule in that it rewrites employees---there is
pattern matching on the left-hand side and term construction on the
right-hand side. The trivial identity monad, \hs{Id}, is used here
because the traversal is a pure function---without even the potential
of failure. \autoref{F:illu-scenarios} also proposed a slightly more
involved transformation scenario for companies: decrease the salaries
of all non-top-level managers. We leave this scenario as an exercise
for the reader.


\begin{figure}[t!]
\begin{boxedminipage}{\hsize}
\begin{lstlisting}[columns=fullflexible,aboveskip=7pt]
-- Queries as generic functions
type Q r = forall x. Term x => x -> r

-- Combinator types
constQ :: r -> Q r
failQ :: MonadPlus m  => Q (m r)
bothQ :: Q u -> Q u' -> Q (u,u')
choiceQ :: MonadPlus m  => Q (m r) -> Q (m r) -> Q (m r)
allQ :: Q r -> Q [r]
adhocQ :: Term x => Q r -> (x -> r) -> Q r

-- Trivial combinator definitions
constQ r = const r
failQ = const mzero
bothQ f g x = (f x, g x)
choiceQ f g x = f x `mplus` g x

-- Non-trivial combinator definitions using pseudo-code
allQ f (C t1 ... tn) = 
    [f t1, ..., f tn]

adhocQ s f x =
    if typeOf x == argumentTypeOf f
       then f x
       else s x
\end{lstlisting}
\end{boxedminipage}
\caption{\textbf{Embedding strategies (queries) in Haskell.}}
\label{F:Q}
\figskip
\end{figure}


\mySubsection{A note on queries}
\label{S:queries}

For most of the paper we focus on transformations since queries do not
seem to add any additional, fundamental challenges. However, we extend
the embedding approach here to include queries for a more complete
illustration of strategic programming. \autoref{F:Q} defines the
generic function type for queries and corresponding function
combinators.

The generic type \hs{Q} models that queries may be applied to terms of
arbitrary types while the result type \hs{r} of the query is a
parameter of the query; it does not depend on the actual type of the
input term. Here, we note that \hs{Q} is not parameterized by a
monad-type constructor, as it is the case for \hs{T}. This design
comes without loss of generality because the result type \hs{r} may be
instantiated also to the application of a monad-type constructor, if
necessary.

The basic strategy \hs{constQ r} denotes the polymorphic constant
function, which returns \hs{r}---no matter the input term or its type.
The basic strategy \hs{failQ} is the always failing query. There is no
special sequential composition for queries because regular function
compostion is appropriate---given that the result of a query is of a
fixed type. However, there is the combinator \hs{bothQ} which applies
two queries to the input and returns both results as a pair. Further,
there is also a form of choice, \hs{choiceQ}, for queries. Ultimately,
there is also one-layer traversal for queries. We only show the
combinator \hs{allQ}, which essentially constructs a list of queried
subterms.  Finally, there is also a form of strategy extension for
queries.


\begin{figure}[t!]
\begin{boxedminipage}{\hsize}
\begin{lstlisting}[columns=fullflexible,aboveskip=7pt]
-- Query each node and collect all results in a list
full_cl :: Monoid u => Q u -> Q u
full_cl s = mconcat . uncurry (:) . bothQ s (allQ (full_cl s))

-- Collection with stop
stop_cl :: Monoid u => Q (Maybe u) -> Q u
stop_cl s = maybe mempty id
          . (s `choiceQ` (Just . mconcat . allQ (stop_cl s)))

-- Find a node to query in top-down, left-to-right manner
once_cl :: MonadPlus m => Q (m u) -> Q (m u)
once_cl s = s `choiceQ` (msum . allQ (once_cl s))
\end{lstlisting}
\end{boxedminipage}
\caption{\textbf{Traversal schemes for queries embedded in Haskell.}}
\label{F:schemes-q}
\figskip
\end{figure}


\begin{figure}[t!]
\begin{boxedminipage}{\hsize}
\small
The \hs{Monoid} type class encapsulates a type with a binary operation, \hs{mappend}, and a unit, \hs{empty}, for that operation:
\begin{lstlisting}
class Monoid a where
  mempty  :: a
  mappend :: a -> a -> a
  mconcat :: [a] -> a
  mconcat = foldr mappend mempty
\end{lstlisting}
The simplest instance is the list monoid, which indeed suggests the names used in the class.
\begin{lstlisting}
instance Monoid [a] where
  mempty  = []
  mappend = (++)
\end{lstlisting}
Other instances are given by addition and zero (or multiplication and one) over numbers, wrapped by the \hs{Sum} constructor: 
\begin{lstlisting}
newtype Sum a = Sum { getSum :: a }     
instance Num a => Monoid (Sum a) where
  mempty = Sum 0
  Sum x `mappend` Sum y = Sum (x + y)
\end{lstlisting}
In each case \hs{mconcat} is used to accumulate a list of values into a single value. This value will be independent of the way in which the accumulation is done if the instance satisfies the \hs{Monoid} laws:
\begin{lstlisting}
mappend mempty x         = x
mappend x mempty         = x
mappend x (mappend y z) = mappend (mappend x y) z
\end{lstlisting}
\end{boxedminipage}
\caption{\textbf{A note on the \hs{Monoid} class in Haskell.}}
\label{F:monoids}
\figskip
\end{figure}


\autoref{F:schemes-q} expresses useful traversal schemes with the
embedding. The first two schemes are parameterized over a monoid to
allow for the collection of data in a general
manner.\hsnote{\autoref{F:monoids} gives a brief overview of monoids
  for those unfamiliar with the concept.} (The postfix ``cl'' hints at
``collection''.) That is, the monoid's type provides the result type
of queries and the monoid's binary operation is used to combine results
from querying many subterms. The third traversal scheme in
\autoref{F:schemes-q} deals with finding a single subterm of interest
as opposed to collecting data from many subterms of interest.


\begin{figure}[t!]
\begin{boxedminipage}{\hsize}
\begin{lstlisting}[columns=fullflexible,aboveskip=7pt]
-- Total the salaries of all employees. 
total_all_salaries :: Q Float
total_all_salaries = getSum . full_cl (adhocQ (constQ mempty) f)
 where
  f (Employee _ s) = Sum s

-- Total the salaries of all employees who are not managers.
total_all_non_managers :: Q Float
total_all_non_managers = getSum . stop_cl type_case
 where
  type_case :: Q (Maybe (Sum Float))
  type_case = adhocQ (adhocQ (constQ Nothing) employee) manager
  employee (Employee _ s) = Just (Sum s)
  manager (Manager _) = Just (Sum 0)
\end{lstlisting}
\end{boxedminipage}
\caption{\textbf{Implementation of two query scenarios from \autoref{F:illu-scenarios}.}}
\label{F:total}
\figskip
\end{figure}


\autoref{F:total} composes traversal for the query scenarios on
companies: total salaries of all employees or non-managers, only. The
implementation of the former is straightforward; perhaps surprisingly,
the implementation of the latter is significantly more involved. The
simple collection scheme \hs{full_cl}\ is inappropriate for totaling all
non-managers because the scheme would reach all nodes
eventually---including the employee subterms that are part of manager
terms, from which salaries must not be extracted though. Hence, a
traversal with `stop' is needed indeed. Further, an always failing
default is needed here---again, in contrast to the simpler case of
totaling all salaries. Finally, the solution depends on the style of
data modeling. That is, the assumed data model distinguishes the types
of managers and employees. Hence, we can use an extra type-specific
case for managers to stop collection at the manager level. Without the
type distinction in the data model, the traversal program would need
to exploit the special \emph{position} of managers within department
terms.

The transformation scenario for decreasing the salaries of all
non-top-level managers, which we left as an exercise for the reader,
calls for similarly involved considerations. These illustrations may
help to confirm that programming errors are quite conceivable in
strategic programming---despite the conciseness of the programming
style.


\mySection{Inventory of strategic programming errors}
\label{S:errors}

The implementation of a strategic programming (sub-) problem (say, a
traversal problem) is normally centered around some
\emph{problem-specific ingredients} (`rewrite rules') that have to be
organized in a more or less complex strategy. There are various
decisions to be made and accordingly, there are opportunities for
misunderstanding and programming errors. This section presents a
fine-grained inventory of programming errors by reflecting
systematically on the use of basic strategy combinators and library
abstractions in the implementation of traversal problems. We use a
deceptively simple scenario as the running example. We begin with a
short proposal of the assumed process of designing and implementing
traversal programs, which in itself may improve understanding of
strategic programming and help reducing programming errors. The
following discussion is biased towards transformations, but coverage
of queries would require only a few, simple adaptations.


\mySubsection{Design of traversal programs}
\label{S:design}

Traversal programming is based on the central notion of \emph{terms of
  interest}---these are the terms to be affected by a
transformation. When \emph{designing a traversal}, the terms of
interest are to be identified along several axes:

\begin{description}

\item[Types] Most obviously, terms of interest are of certain
  types. For instance, a transformation for salary increase may be
  concerned with the type of employees.

\smallskip

\item[Patterns and conditions] Terms of interest often need to match
  certain patterns. For instance, a transformation for the application
  of a distributive law deals with the pattern $x*(y+z)$. In addition,
  the applicability of transformations is often subject to
  (pre-) conditions.

\smallskip

\item[Position-based selection] A selection criterion may be applied
  if the transformation should not affect all terms with fitting
  patterns and conditions. In an extreme case, a single term is to be
  selected. Such selection typically refers to the position of these
  terms; think of top-most versus bottom-most.

\smallskip

\item[Origin] The initial input is expected to contribute terms with
  fitting patterns and conditions, but previous applications of
  rewrite rules may contribute terms as well. Hence, it must be
  decided whether the latter kind of origin also qualifies for terms
  of interest. For instance, an unfolding transformation for recursive
  function definitions may specifically disregard function
  applications that were introduced by unfolding.

\end{description}


\mySubsection{Implementation of traversal programs}
\label{S:impl}

When \emph{implementing a traversal}, the types and patterns of terms
of interest are modeled by the left-hand sides of rewrite
rules. Conditions are typically modeled by the rewrite rules, too, but
the choice of the traversal scheme may be essential for being able to
correctly check the conditions. For instance, the traversal scheme may
need to pass down information that is needed by a condition
eventually. The axes of selection and origin (of terms of interest)
are expressed through the choice of a suitable traversal scheme.

Let us provide a summary of basic variation points in traversal
implementation. For simplicity, let us focus here on traversal
problems whose implementation corresponds to a strategy that has been
built by applying one or more traversal schemes from a
library to the problem-specific rewrite rules, possibly subject to
composition of rewrite rules or sub-traversals.

Organizing the strategy involves the following decisions:

{\small\begin{description}
\item[Scheme] Which traversal scheme is to be used?
\begin{itemize}
\item Is a full or a limited traversal required?
\item Does top-down versus bottom-up order matter?
\item Does a strategy need to be iterated?
\item ...
\end{itemize}
\item[Default] What polymorphic and monomorphic defaults are to be used?
\begin{itemize}
\item The identity transformation.
\item The always failing transformation.
\item Another, more specific, behavior.
\end{itemize}
\item[Composition] How to compose a strategy from multiple parts?
\begin{itemize}
\item Use type case (strategy extension) at the level of rewrite rules.
\item Combine arguments of a traversal scheme in a sequence.
\item Combine arguments of a traversal scheme in a choice.
\item Combine traversals in a sequence.
\item Combine traversals in a choice.
\end{itemize}
\end{description}}

Based on an appropriate running example we shall exercise these
choices, and see the consequences of incorrect decisions as they cause
programming errors in practice. In our experience, wrong choices are
the result of insufficiently understanding i) the variation points of
traversal schemes, ii) the subtleties of control flow,
and iii) the axes of terms of interest in practical
scenarios.


\mySubsection{The running example}
\label{S:example}

To use a purposely simple example, consider the transformation problem
of `incrementing all numbers in a term'. Suppose $\ell$ is the rewrite
rule that maps any given number $n$ to $n+1$. It remains to compose a
strategy that can iterate $\ell$ over any term. 

For concreteness' sake, we operate on n-ary trees of natural
numbers. Further, we assume a Peano-like definition of the data type
for numbers. Here are the data types for numbers and trees:

\begin{lstlisting}
data Nat = Zero | Succ Nat
data Tree a = Node {rootLabel :: a, subForest :: [Tree a]}
\end{lstlisting}

The Peano-induced recursion implies a simple form of nesting. It goes
without saying that the Peano-induced nesting form is contrived, but
its inclusion allows us to cover nesting as such---any practical
scenario of traversal programming involves nesting at the
data-modeling level---think of nesting of departments in the company
example, or nesting of expressions or function-definition blocks in
the AST example given in the introduction.

Here are simple tree samples:

\begin{lstlisting}[basicstyle=\footnotesize\rmfamily\itshape]
tree1 = Node { rootLabel = Zero, subForest = [] }  -- A tree of numbers
tree2 = Node { rootLabel = True, subForest = [] }  -- A tree of Booleans
tree3 = Node { rootLabel = Succ Zero, subForest = [tree1,tree1] }  -- Two subtrees
\end{lstlisting}

The rewrite rule for incrementing numbers is represented as follows:

\begin{lstlisting}
increment n = Succ n
\end{lstlisting}

In fact, let us use monadic style because the basic Strafunski-like
library introduced in \S\ref{S:embedding} assumes monadic style for
all combinators---in particular, for all arguments. Hence, we commit
to the \hs{Maybe}\ monad and its constructor \hs{Just}:

\begin{lstlisting}
increment n = Just (Succ n)
\end{lstlisting}

It remains to complete the rewrite rule into a traversal strategy that
increments all numbers in an arbitrary term. That is, we need to make
decisions regarding traversal scheme, default, and composition for the
implementation. 

Given the options \hs{full_td}, \hs{full_bu}, \hs{stop_td},
\hs{once_bu}, \hs{once_td}, and \hs{innermost}, which traversal scheme
is the correct one for the problem at hand? Also, how to exactly apply
the chosen scheme to the given rewrite rule? An experienced strategist
may quickly exclude a few options. For instance, it may be obvious
that the scheme \hs{once_bu}\ is not appropriate because we want to
increment \emph{all} numbers, while \hs{once_bu}\ would only affect
one number. In the remainder of the section, we will attempt different
schemes and vary other details, thereby showcasing potential
programming errors.


\mySubsection{Strategies going wrong}

The composed strategy may go wrong in different ways:
\begin{itemize}
\item It diverges.
\item It transforms incorrectly, i.e., numbers are not exactly incremented.
\item It does not modify the input, i.e., numbers are not incremented at all.
\item It fails even when the transformation is assumed never to fail.
\item It succeeds even when failure is preferred for terms without numbers.
\end{itemize}
Let us consider specific instances of such problems.


\mySubsubsection{Divergent traversal}
\label{S:divergence}

Let us attempt a full top-down traversal. Alas, the traversal
diverges:\hsnote{Throughout the section, we operate at the Haskell
  prompt. That is, we show input past the `$>$' prompt sign and
  resulting output, if any, right below the input.}

\begin{lstlisting}[columns=fullflexible]
> full_td (adhocT idT increment) tree1
... an infinite tree is printed ...
\end{lstlisting}

The intuitive reason for non-termination is that numbers are
incremented prior to the traversal's descent. Hence, the term
under traversal grows and each increment enables another
increment.

Let us attempt instead the \hs{innermost} scheme. Again, traversal
diverges:

\begin{lstlisting}[columns=fullflexible]
> innermost (adhocT failT increment) tree1
... no output ever is printed ...
\end{lstlisting}

The combinator \hs{innermost}\ repeats \hs{once_bu}\ until it fails,
but it never fails because there is always a redex to which to apply
the \hs{increment}\ rule. Hence, \hs{tree1}\ is rewritten
indefinitely. 

Both decisions here illustrate the case of choosing the wrong
traversal scheme which in turn may be the result of insufficiently
understanding some axes of terms of interest (see \S\ref{S:design})
and associated properties of rewrite rules. In particular, the
traversal schemes used here support the \emph{origin} axis in a way
terms of interest are created by the traversal.


\mySubsubsection{Incorrect transformation}
\label{S:incorrectx}

Let us attempt instead the \hs{full_bu} scheme:

\begin{lstlisting}[columns=fullflexible]
> full_bu (adhocT idT increment) tree1
Just (Node {rootLabel = Succ Zero, subForest = []})
\end{lstlisting}

The root label was indeed incremented. This particular test case looks
fine, but if we were testing the same strategy with trees that contain
non-zero numbers, then we would learn that the composed strategy
replaces each number $n$ by $2n+1$ as opposed to $n+1$. To see this,
one should notice that a number $n$ is represented as a term of depth
$n+1$, and the choice of the scheme \hs{full_bu}\ implies that
\hs{increment}\ applies to each `sub-number'. 

More generally, we see an instance of an overlooked
\emph{applicability condition} (see \S\ref{S:design}) in that numbers
are terms of interest, but not subterms thereof. The same kind of
error could occur in the implementation of any other scenario as long
as it involves nesting. In real-world scenarios, nesting may
actually arise also through mutual (data type-level) recursion.


\mySubsubsection{No-op traversal}
\label{S:noop}

Finally, let us attempt the \hs{stop_td} scheme. Alas, no incrementing happens:

\begin{lstlisting}[columns=fullflexible]
> stop_td (adhocT idT increment) tree1
Just (Node {rootLabel = Zero, subForest = []})
\end{lstlisting}

That is, the result equals \hs{Just tree1}. The problem is that the
strategy should continue to descend as long as no number is hit, but
the polymorphic default \hs{idT}\ makes the strategy stop for any
subterm that is not a number. Let us replace \hs{idT}\ by
\hs{failT}. Finally, we arrive at a proper solution for the original
problem statement:

\begin{lstlisting}[columns=fullflexible]
> stop_td (adhocT failT increment) tree1
Just (Node {rootLabel = Succ Zero, subForest = []})
\end{lstlisting}

Hence, \hs{stop_td} is the correct traversal scheme for the problem at
hand, but we also need to be careful about using the correct
polymorphic default for lifting the rewrite rule to the strategy
level; see \S\ref{S:impl}.


\mySubsubsection{Unexpected failure}
\label{S:unexpected-fail}

\hs{failT}\ is the archetypal polymorphic default for certain schemes,
while it is patently inappropriate for others. To see this, suppose,
we indeed want to replace each number $n$ by $2n+1$, as we
accidentally ended up doing in \S\ref{S:incorrectx}. Back then, the
polymorphic default \hs{idT}\ was appropriate for \hs{full_bu}.  In
contrast, the default \hs{failT}\ is not appropriate:

\begin{lstlisting}[columns=fullflexible]
> full_bu (adhocT failT increment) tree1
Nothing
\end{lstlisting}

This is a case of unexpected failure in the sense that we expect the
traversal for incrementing numbers to succeed for all possible input
terms. The problem is again due to the wrong choice of default.


\mySubsubsection{Unexpected success}
\label{S:unexpected-success}

Let us apply the confirmed scheme and default to a tree of Booleans:

\begin{lstlisting}[columns=fullflexible]
> stop_td (adhocT failT increment) tree2
Just (Node { rootLabel = True, subForest = [] })
\end{lstlisting}

Of course, no incrementing happens; the output equals the
input. Arguably, a strategic programmer could expect that the
traversal should fail, if the rewrite rule for incrementing never
applies. For comparison, the traversal scheme \hs{once_bu} does indeed
fail in case of inapplicability of its argument. Defining a suitable
variation on \hs{stop_td} that indeed fails in the assumed way we
leave as an exercise for the reader.  Misunderstood success and
failure behavior may propagate as a programming error as it may affect
the control flow in the strategic program.


\mySubsection{Subtle control flow}
\label{S:cflow}

Arguably, several of the  problems discussed involve subtleties of
control flow. It appears to be particularly difficult to understand
and to correctly configure control flow of strategies on the grounds
of success and failure behavior for operands in strategy composition.

Let us modify the running example slightly to provide another
illustration. We consider the refined problem statement that only \emph{even}
numbers are to be incremented. In the terminology of rewriting,
this statement calls for a conditional rewrite rule:\footnote{Both the
  original \hs{increment}\ function and the new `conditional'
  \hs{increment_even}\ function go arguably beyond the basic notion of
  a rewrite rule that requires a non-variable pattern on the left-hand
  side. We could easily recover classic style by using two rewrite
  rules---one for each form of a natural number.}

\medskip

\begin{lstlisting}
-- Pseudo code for a conditional rewrite rule
increment_even : n -> Succ(n) where even(n)

-- Haskell code (monadic notation)
increment_even n = do guard (even n); increment n
\end{lstlisting}

We use the same traversal scheme as before:

\begin{lstlisting}[columns=fullflexible]
> stop_td (adhocT failT increment_even) tree1
Just (Node {rootLabel = Succ Zero, subForest = []})
\end{lstlisting}

This particular test case looks fine, but if we were testing the same
strategy with trees that contain odd numbers, then we would learn
that the composed strategy in fact also increments those. The problem
is that the failure of the precondition for \hs{increment}\ propagates
to the traversal scheme which takes failure to mean `continue
descent'. However, once we descend into odd numbers, we will hit an
even sub-number in the next step, which is hence incremented. So we
need to make sure that recursion ceases for \emph{all} numbers. Thus:
 
\begin{lstlisting}
increment_even n    | even n      = Just (Succ n)
                    | otherwise   = Just n
\end{lstlisting}

The example shows the subtleties of control flow in strategic
programming: committing to the specific monomorphic type in
\hs{adhocT} can still fail, and so lead to further traversal.


\mySubsection{Dead code}
\label{S:dead-code}

So far we have mainly spotted programming errors through comparison of
expected with actual output, if any. Let us now switch to the
examination of composed strategies.  There are recurring patterns of
producing dead code in strategic programming. We take the position
here that dead code is a symptom of programming errors.

Consider the following patterns of strategy expressions:

\begin{itemize}
\item \hs{adhocT}~(\hs{adhocT}~$s$~$f_1$)~$f_2$
\item \hs{choiceT}~$s_1$~$s_2$
\item \hs{sequT}~$s_1$~$s_2$
\end{itemize}

In the first pattern, if the operands $f_1$ and $f_2$ are of the
same type (or more generally, the type of $f_2$ can be specialized to
the type of $f_1$), then $f_1$ has no chance of being
applied. Likewise, in the second pattern, if $s_1$ never possibly
fails, then $s_2$ has no chance of being applied.  Finally, in the
third pattern, if $s_1$ never possibly succeeds, which is likely to be
the symptom of a programming error by itself, then, additionally,
$s_2$ has no chance of being applied.

Let us illustrate the first kind of programming error: two
type-specific cases of the same type that are composed with
\hs{adhocT}. Let us consider a refined problem statement such that
incrementing of numbers is to be replaced by (i) increment \emph{by
  one}  all odd numbers, (ii) increment \emph{by two}  all even
numbers. Here are the basic building blocks that we need:

\begin{lstlisting}
atOdd n     | odd n        = Just (Succ n)
            | otherwise    = Nothing
                         
atEven n    | even n       = Just (Succ (Succ n))
            | otherwise    = Nothing
\end{lstlisting}

Arguably, both rewrite rules could also have been combined in a single
function to start with, but we assume here a modular decomposition as
the starting point. We also leave it as an exercise to the reader to
argue that the monomorphic default \hs{Nothing}\ is appropriate for
the given problem. Intuitively, we wish to compose these type-specific
cases so that both of them are tried.

Let us attempt a composition that uses \hs{adhocT} twice:

\begin{lstlisting}[columns=fullflexible]
> stop_td (adhocT (adhocT failT atEven) atOdd) tree1
Just (Node {rootLabel = Zero, subForest = []})
\end{lstlisting}

Alas, no incrementing seems to happen. The problem is that there are
two type-specific cases for numbers, and the case for odd numbers
dominates the one for even numbers. The case for even numbers is
effectively dead code. In the sample tree, the number, \hs{Zero}, is
even.

The two rewrite rules need to be composed at the monomorphic level of
the number type---as opposed to the polymorphic level of strategy
extension. To this end, we need composition combinators that can be
applied to functions of specific types as opposed to generic
functions:

\begin{lstlisting}
msequ :: Monad m => (x -> m x) -> (x -> m x) -> x -> m x
msequ s s' x = s x >>= s'

mchoice :: MonadPlus m => (x -> m x) -> (x -> m x) -> x -> m x
mchoice f g x = mplus (f x) (g x)
\end{lstlisting}

Using \hs{mchoice}, we arrive at a correct composition:

\begin{lstlisting}[columns=fullflexible]
> stop_td (adhocT failT (mchoice atEven atOdd)) tree1
Just (Node {rootLabel = Succ (Succ Zero), subForest = []})
\end{lstlisting}

We face a more subtle form of dead code when the root type for terms
in a traversal implies that the traversal cannot encounter subterms of
the type expected by a type-specific case. Consider again the strategy
application that we already used for the illustration of potentially
unexpected success in \S\ref{S:unexpected-success}:

\begin{lstlisting}[columns=fullflexible]
> stop_td (adhocT failT increment) tree2
Just (Node { rootLabel = True, subForest = [] })
\end{lstlisting}

The output equals the input. In this application, the rewrite rule
\hs{increment} is effectively dead code. In fact, it is not important
what actual input is passed to the strategy. It suffices to know that
the input's type is \hs{Tree}~\hs{Boolean}. Terms of interest, i.e.,
numbers, cannot possibly be found below any root of type
\hs{Tree}~\hs{Boolean} and the given strategy is a no-op in such a
case. This may be indeed a symptom of a programming error: we either
meant to traverse a different term (i.e., one that contains numbers)
or we meant to invoke a different strategy (i.e., one that affects
Boolean literals or polymorphic trees). Accordingly, one could argue
that the strategy application at hand should be rejected statically.


\mySubsection{Options of composition}
\label{S:composition}

As a final exercise on the matter of strategy composition, let us
study one more time the refined example for incrementing odd and even
numbers as introduced in \S\ref{S:dead-code}. We take for granted the
following decisions: \hs{stop_td} for the traversal scheme and
\hs{failT} for the polymorphic default. Given all the principal
options for composition, as of \S\ref{S:impl}, there are the following
concrete options for the example:

\medskip

{\footnotesize\begin{enumerate}[1.]
\item \hs{stop_td (adhocT (adhocT failT atEven) atOdd)}
\item \hs{stop_td (adhocT failT (mchoice atEven atOdd))}
\item \hs{stop_td (adhocT failT (msequ atEven atOdd))}
\item \hs{stop_td (choiceT (adhocT failT atEven) (adhocT failT atOdd))}
\item \hs{stop_td (sequT (adhocT failT atEven) (adhocT failT atOdd))}
\item \hs{choiceT (stop_td (adhocT failT atEven)) (stop_td (adhocT failT atOdd))}
\item \hs{sequT (stop_td (adhocT failT atEven)) (stop_td (adhocT failT atOdd))}
\end{enumerate}}

\medskip

(We do not exercise all variations on the order of operands.) Option
(1.) had been dismissed already because the two branches involved are
of the same type. Option (2.) had been approved as a correct
solution. Option (4.) turns out to be equivalent to option (2.). (This
equivalence is implied by basic properties of defaults and composition
operators.) The strategies of the other options do not implement the
intended operation. Demonstrating and explaining the issues with these
strategies we leave as an exercise for the reader.


\mySection{Static typing of traversal strategies}
\label{S:typing}

We use established means of static typing to curb the identified
programming errors, to the extent possible, in a way that basic
strategy combinators and library abstractions are restricted in
generality. In particular, we use static typing to avoid wrong
decisions regarding strategy composition, to reduce subtleties of
control flow, and to avoid some forms of dead code. We do not design
new type systems here. Instead, we attempt to leverage established
means, as well as we can.

The section is organized as a sequel of contributions---each of them
consisting of language-agnostic advice for improving strategic
programming and an illustration in Haskell.
We use Haskell for illustrations because it is an established
programming language for statically typed strategic programming and
its type system is rather powerful in terms of supporting different
forms of polymorphism and simple forms of dependent
typing~\cite{McBride02,Thiemann02,Shan04,KiselyovLS04-HW,RodriguezJJGKO08}.
A basic `reading knowledge' of Haskell, supplemented with the
background notes in \S\ref{S:background}, should be sufficient to
understand the essence of the Haskell illustrations.

We provide language-agnostic advice because different languages may
need to achieve the suggested improvements in different ways, if at
all. In fact, not even Haskell's advanced type system achieves the
suggested improvements in a fully satisfactory manner. Hence, the
section may ultimately suggest improvements of practical type systems
or appropriate use of existing proposals for type-system improvements, e.g.,
\cite{CraryW99,ShieldsM01,Epigram,JonesVWS07,Leijen08,XuJC09,Agda,OliveiraMO10}.


\newcommand{\adviceDefault}{Hard-wire defaults into traversal schemes}
\mySubsection{\adviceDefault}
\label{S:adviceDefault}

\begin{advice} 
\label{A:default}
By  hard-wiring a suitable
default into each traversal scheme, rule out wrong decisions regarding the polymorphic default during
strategy composition (see \S\ref{S:impl}). Here we assume that the default
can either be statically defined for each scheme or else that it can be determined at
runtime by observing other arguments of the scheme.
\end{advice}

We can illustrate the advice in Haskell in a specific manner by
reducing the polymorphism of traversal schemes as follows. While the
general schemes of \S\ref{S:embedding} are essentially parameterized
by polymorphic functions on terms (in fact, rank-2 polymorphic
functions~\cite{LaemmelPJ03,JonesVWS07}), the restricted schemes are
parameterized by specific type-specific cases. There is also a
proposal for a variation of `Scrap Your Boilerplate' that points in
this direction~\cite{MitchellR07}.

The following primed definitions take a type-specific case \hs{f},
which is then generalized \emph{within the definition} by means of the
appropriate polymorphic default, \hs{idT}{} or \hs{failT}. We delegate
to the more polymorphic schemes otherwise.

\begin{lstlisting}[columns=fullflexible]
full_td' :: (Term x, Term y, Monad m) => (x -> m x) -> y -> m y
once_bu' :: (Term x, Term y, MonadPlus m) => (x -> m x) -> y -> m y
stop_td' :: (Term x, Term y, MonadPlus m) => (x -> m x) -> y -> m y
...
\end{lstlisting}

\begin{lstlisting}[columns=fullflexible]
full_td' f = full_td (adhocT idT f)
once_bu' f = once_td (adhocT failT f)
stop_td' f = stop_td (adhocT failT f)
...
\end{lstlisting}

These schemes reduce programming errors as follows. Most obviously,
polymorphic defaults are correct by design because they are hard-wired
into the definitions. A side effect is that the use of strategy
extension is now limited to the library, and hence
strategy composition is made simpler by reducing the number of options (see
\S\ref{S:composition}).

However, there are scenarios that require the general schemes; see
\cite{VisserBT98,LaemmelV02-PADL,LaemmelV03-PADL} for examples. The
problem is that we may need a variable number of type-specific cases.
Some scenarios of strategies with multiple cases can be decomposed
into multiple traversals, but even when it is possible, it may be
burdensome and negatively affect performance. Further, there are cases,
when the hard-wired default is not applicable. Hence, the default
should be admissible to overriding.

\newcommand{\twoLibs}{As a result, the restricted schemes cannot fully
  replace the general schemes. Therefore, a strategic programming
  library would need to provide both variants and stipulate usage of
  the restricted schemes whenever possible.}

\twoLibs

In principle, one can think of unified schemes that can be applied to
single type-specific cases, collections thereof, and polymorphic
functions that readily incorporate a polymorphic default. Those schemes
would need to coerce type-specific cases to polymorphic functions. We
will illustrate this idea in a limited manner in
\S\ref{S:adviceTypecase}.


\newcommand{\adviceFallibility}{Declare and check fallibility contracts}
\mySubsection{\adviceFallibility}
\label{S:adviceFallibility}

\begin{advice}
\label{A:fallibility}
Curb programming errors due to subtle control flow (see
\S\ref{S:cflow}) by declaring and checking contracts regarding
fallibility. These contracts convey whether the argument of a possibly
restricted traversal scheme is supposed to be fallible and whether the
resulting traversal is guaranteed to be infallible (subject to certain
preconditions).
\end{advice}

The advice is meant to improve strategic programming so that more
guidance is provided as far as the success and failure behavior of
traversal schemes and their arguments is concerned. According to
\S\ref{S:properties}, traversal schemes differ in terms of their
fallibility properties and the dependence of these properties on
fallibility properties of the arguments. For instance, \hs{full_td}
preserve infallibility, that is, a composed traversal \hs{full_td s}
is infallible if the argument \hs{s} is infallible. In contrast,
\hs{stop_td s} is infallible regardless of \hs{s}.

The function signatures of the schemes of \S\ref{S:embedding} hint at
fallibility properties: see the distinguished use of
\hs{Monad} vs.\ \hs{MonadPlus}. For instance:

\begin{lstlisting}[columns=fullflexible]
full_td :: Monad m => T m -> T m
once_bu :: MonadPlus m => T m -> T m
stop_td :: MonadPlus m => T m -> T m
\end{lstlisting}

However, such hinting does not imply checks.  For instance, a
programmer may still pass a notoriously failing argument to
\hs{full_td} despite the signature's hint that a universally
succeeding argument may be perfectly acceptable.  Such hinting may
also be misleading. For instance, the appearance of \hs{MonadPlus} in
the type of \hs{stop_td} may suggest that such a traversal may fail,
but, in fact, it cannot.  Instead, the appearance of \hs{MonadPlus}
hints at the fact that the argument is supposed to be fallible.

We can illustrate the advice in Haskell in a specific manner by
providing infallible variations on the traversal schemes of
\S\ref{S:embedding}. To this end, we use the identity monad whenever
we want to require or imply infallibility. We use the maybe monad
whenever the argument of such a scheme is supposed to be
fallible. Thus:

\begin{lstlisting}[columns=fullflexible]
full_td' :: T Id -> T Id
full_bu' :: T Id -> T Id
stop_td' :: T Maybe -> T Id
innermost' :: T Maybe -> T Id
repeat' :: T Maybe -> T Id
try' :: T Maybe -> T Id
\end{lstlisting}

Applications of these restricted schemes are hence guaranteed to be
infallible. We cannot provide an infallible variation on \hs{once_bu}
due to its nature. It is also instructive to notice that \hs{try}
models transition from a fallible to an infallible strategy---not just
operationally, as before, but now also at the type level. The inverse
transition is not served.

The primed definitions \hs{full_td'}~and \hs{full_bu'}~simply delegate
to the original schemes, but the other primed definitions need to be
defined from scratch because they need to compose infallible and
fallible strategy types in a manner that requires designated forms of
sequence and choice. These new definitions can be viewed as
constructive proofs for fallibility properties.

\begin{lstlisting}[columns=fullflexible]
full_td' s    = full_td s
full_bu' s    = full_bu s
stop_td' s    = s `choiceT'` allT (stop_td' s)
innermost' s  = repeat' (once_bu s)
repeat' s     = try' (s `sequT'` repeat' s)
try' s        = s `choiceT'` idT

choiceT' :: T Maybe -> T Id -> T Id
choiceT' f g x = maybe (g x) Id (f x)

sequT' :: T Maybe -> T Id -> T Maybe
sequT' f g = f `sequT` (Just . getId . g)
\end{lstlisting}

The type of \hs{choiceT'} is interesting in so far that it allows us
to compose a fallible strategy with an infallible strategy to obtain
an infallible strategy. That is, the scope of fallibility is made local.

The fallibility properties were modeled at the expense of eliminating
the general monad parameter. Generality could be recovered though by
consistently parameterizing all infallible schemes with a plain monad
and adding an application of the monad transformer for \hs{Maybe} whenever
the argument of such a scheme is supposed to be fallible. Thus:

\begin{lstlisting}[columns=fullflexible]
full_td'' :: Monad m => T m -> T m -- equals original type
full_bu'' :: Monad m => T m -> T m -- equals original type
stop_td''   :: Monad m => T (MaybeT m) -> T m
innermost'' :: Monad m => T (MaybeT m) -> T m
repeat''    :: Monad m => T (MaybeT m) -> T m
try''       :: Monad m => T (MaybeT m) -> T m
\end{lstlisting}

The definitions are omitted here as they require non-trivial knowledge
of monad transformers; see though the paper's online code
distribution. These definitions declare the fallibility contracts
better than the original schemes, but enforcement is limited. The
monad-type parameter may be still (accidentally) instantiated to an
instance of \hs{MonadPlus}. For instance, the types of \hs{full_td''}
and \hs{full_bu''} are not at all constrained, when compared to the
original schemes.


\newcommand{\adviceCflow}{Reserve fallibility for modeling control flow}
\mySubsection{\adviceCflow}
\label{S:adviceCflow}

\begin{advice} 
\label{A:cflow}
Curb programming errors due to subtle control flow (see
\S\ref{S:cflow}) by reserving fallibility, as discussed so far, for
modeling control flow. If success and failure behavior is needed for
other purposes, such as assertion checking, then strategies shall use
effects that cannot be confused with efforts to model control
flow. The type system must effectively rule out such confusion.
\end{advice}

We can illustrate the advice in Haskell in a specific manner by defining
the traversal schemes of \S\ref{S:embedding} from scratch in terms of 
two distinct types for infallible versus fallible types: 

\begin{lstlisting}[columns=fullflexible]
data T m = T { getT :: forall x. Term x => x -> m x }
data T' m = T' { getT' :: forall x. Term x => x -> MaybeT m x }
\end{lstlisting}

We use data types here (as opposed to type synonyms) so that the two
types cannot possibly be confused. This is discussed in more detail
below.

If an infallible strategy needs to fail for reasons other than
affecting regular control flow, then the monad parameter can still be
used to incorporate the maybe monad or an exception monad, for
example. In this manner, strategies may perform assertion checking, as
often needed for preconditions of nontrivial transformations, without
running a risk of failure to be consumed by the control-flow semantics
of the strategic program. In this manner, strategic programming is
updated to reliably separate control flow and exceptions (or other
effects), as it is common in the general programming
field~\cite{RyderS03,Marlow06}.

The types of the `full' traversal schemes reflect that control flow is
hard-wired:

\begin{lstlisting}[columns=fullflexible]
full_td   :: Monad m => T m -> T m
full_bu   :: Monad m => T m -> T m
\end{lstlisting}

The types of the `once' traversal schemes reflect that fallibility is essential:

\begin{lstlisting}[columns=fullflexible]
once_td   :: Monad m => T' m -> T' m
once_bu   :: Monad m => T' m -> T' m
\end{lstlisting}

The other library combinators construct infallible strategies from
fallible ones:
 
\begin{lstlisting}[columns=fullflexible]
stop_td   :: Monad m => T' m -> T m
innermost :: Monad m => T' m -> T m
repeat    :: Monad m => T' m -> T m
try       :: Monad m => T' m -> T m
\end{lstlisting}

At first sight, these types look deceptively similar to those that we
defined for fallibility contracts in
\S\ref{S:adviceFallibility}. However, the important difference is that
\hs{T m} and \hs{T' m'} cannot be confused whereas this is possible
for \hs{T m} and \hs{T (MaybeT m')} if \hs{m} and \hs{MaybeT m'} are
unifiable. 

The definitions of the new schemes are omitted here as they rely on a
designated, non-trivial suite of basic strategy combinators; see
though the paper's online code distribution. It is fair to say that
the present illustration also addresses Advice~\ref{A:fallibility}
regarding fallibility contracts.


\newcommand{\adviceTypecase}{Enable families of type-specific cases}
\mySubsection{\adviceTypecase}
\label{S:adviceTypecase}

\begin{advice}
\label{A:typecase}
Rule out dead code due to overlapping type-specific cases (see
\S\ref{S:dead-code}) by enabling strongly typed families of
type-specific cases as arguments of traversal schemes. Such a family
is a non-empty list of functions the types of which are pairwise
non-unifiable but they all instantiate the same generic function type
for strategies.
\end{advice}

We can illustrate the advice in Haskell in a specific manner by making
use of advanced type-class-based programming. More specifically, we
leverage existing library support for strongly typed, heterogenous
collections---the HList library~\cite{KiselyovLS04-HW}.


\begin{figure}[t!]
\begin{boxedminipage}{\hsize}
\small
\begin{lstlisting}[columns=fullflexible,aboveskip=2pt]
-- Type-class-polymorphic type of familyT
class (Monad m, HTypeIndexed f) => FamilyT f m
  where
    familyT :: T m -> f -> T m

-- Empty list case
instance Monad m => FamilyT HNil m
  where 
    familyT g _ = g

-- Non-empty list case
instance ( Monad m
          , FamilyT t m
          , Term x
          , HOccursNot (x -> m x) t
          )
            => FamilyT (HCons (x -> m x) t) m
  where 
    familyT g (HCons h t) =  adhocT (familyT g t) h
\end{lstlisting}

The list of type-specific cases is constrained to only hold elements
of distinct types; see the constraint \hs{HTypeIndexed}, which is
provided by the HList library. Also notice that the element types are
constrained to be function types for monadic transformations; see the
pattern \hs{x -> m x} in the head of the last instance.  As a proof
obligation for the \hs{HTypeIndexed}{} constraint, the instance for
non-empty lists must establish that the head's type does not occur
again in the tail of the family; see the constraint \hs{HOccursNot},
which is again provided by the HList library.
\end{boxedminipage}
\caption{Derivation of a transformation from type-specific cases and a
  default.}
\label{F:familyT}
\figskip
\end{figure}


Consider the pattern of composing a strategy from several (say, two)
type-specific cases and a polymorphic default:

\begin{quote}
\hs{adhocT (adhocT} $s$ $f_1$\hs{)} $f_2$
\end{quote}

The type-specific cases, $f_1$ and $f_2$, are supposed to override the
polymorphic default $s$ in a point-wise manner. Ignoring static typing
for a second, we can represent the two type-specific cases instead as
a list $[f_1, f_2]$. Conceptually, it is a heterogenous list in the
sense that the types of the functions for type-specific cases are
supposed to be distinct (in fact, non-unifiable); otherwise dead code
is admitted. Using HList's constructors for heterogenous lists, the
type-specific cases are represented as follows:

\begin{quote}
\hs{HCons} $f_1$ \hs{(HCons} $f_2$ \hs{HNil)}
\end{quote}

Now suppose that the types of the type-specific cases all instantiate
the polymorphic type \hs{T}. Further suppose that there is a function
\hs{familyT} for strategy construction; it takes two arguments: the
heterogenous collection and a transformation $s$ which serves as
polymorphic default. The original strategy is constructed as follows:

\begin{quote}
\hs{familyT (HCons} $f_1$ \hs{(HCons} $f_2$ \hs{HNil))} $s$
\end{quote}

The function \hs{familyT} must be polymorphic in a special manner such
that it can process all heterogenous collections of type-specific
cases. To this end, the function must be overloaded on all possible
types of such collections, which is achieved by type-class-based
programming; see \autoref{F:familyT}.

The family-enabled traversal schemes are defined as follows:

\begin{lstlisting}[columns=fullflexible]
full_td' s     = full_td (familyT idT s)
full_bu' s     = full_bu (familyT idT s)
once_td' s     = once_td (familyT failT s)
once_bu' s     = once_bu (familyT failT s)
stop_td' s     = stop_td (familyT failT s)
innermost' s   = innermost (familyT failT s)
\end{lstlisting}

That is, the family-enabled schemes invoke the \hs{familyT} function
to resolve the heterogenous collection into a regular generic function
subject to the polymorphic default that is known for each traversal
scheme. In this manner, we do not just avoid dead code for
type-specific cases; we also address the issue of
Advice~\ref{A:default} in that the polymorphic default is hard-wired
into the schemes---though without the restriction to a single
type-specific case, as was the case in \S\ref{S:adviceDefault}.

Admittedly, the illustration involves substantial encoding.  For
example, type errors in type-class-based programming are rather
involved, and tend to be couched in terms well below the abstraction
level of the strategic programmer. Hence, future type systems should
provide first-class support for the required form of type case.


\newcommand{\adviceReachability}{Declare and check reachability contracts}
\mySubsection{\adviceReachability}
\label{S:adviceReachability}

\begin{advice}
\label{A:reachability}
Curb programming errors due to type-specific cases not exercised
during traversal (see \S\ref{S:dead-code}) by declaring and checking
contracts regarding reachability. These contracts describe that the
argument types of type-specific cases can possibly be encountered on
subterm positions of the traversal's input whose root type is
known. The type system shall enforce these contracts for composed
traversals.
\end{advice}

We can illustrate the advice in Haskell in a specific manner by making
use again of advanced type-class-based programming. That is, we
leverage a type-level relation on types, which captures whether or not
terms of one type may occur within terms of another type. The relation
is modelled by the following type class:

\begin{lstlisting}[columns=fullflexible]
class ReachableFrom x y
instance ReachableFrom x x -- reflexivity
-- Other instances are derived from data types of interest.
\end{lstlisting}

The above instance makes sure that relation \hs{ReachableFrom} is
reflexive. All remaining instances must be derived from the
declarations of data types that are term types. The instances
essentially rephrase the constructor components of the data-type
declarations. For instance, the data type for polymorphic trees and
the leveraged data type for polymorphic lists imply the following
contributions to the relation \hs{ReachableFrom}:

\begin{lstlisting}[columns=fullflexible]
instance ReachableFrom a [a]
instance ReachableFrom a (Tree a)
instance ReachableFrom [Tree a] (Tree a)
\end{lstlisting}

Recall the example of dead code due to unreachable types in
\S\ref{S:dead-code}. The relation \hs{ReachableFrom}    clearly demonstrates that numbers can
be reached from a tree of numbers (using the second instance) but not
from a tree of Booleans.

The relation \hs{ReachableFrom} can be leveraged directly for the
declaration of contracts regarding reachability. To this end,
appropriate constraints must be added to the function signatures of
the traversal schemes. For simplicity, we focus here on the simpler
function signatures for the traversal schemes of
\S\ref{S:adviceDefault} with hard-wired defaults. Thus:

\begin{lstlisting}[columns=fullflexible]
full_td'' :: (Term x, Term y, Monad m, ReachableFrom x y)
          => (x -> m x)
          -> y -> m y
full_td'' = full_td'
\end{lstlisting}

Compared to \hs{full_td'}~of \S\ref{S:adviceDefault}, the
\hs{ReachableFrom} constraint has been added. To this end, the
application of the forall-quantified type synonym for the resulting
strategy had to be inlined so that the constraint can address the type
variable \hs{y} for the strategy type. Of course, the constraint does
not change the behavior of full top-down traversal, and hence, there
is no correctness issue. However, it is not straightforward to see or
to prove that the additional constraint does not remove any useful
behavior. Here, we consider the deep identity traversal or the
completely undefined traversal as `useless'.

While the explicit declaration of contracts provides valuable
documentation, it is possible, in principle, to infer reachability
constraints from the traversal programs themselves. This may
be difficult with type-class-based programming, but we consider
a corresponding static program analysis in \S\ref{S:adviceProveReachability}.

The idea underlying this illustration has also been presented in
\cite{Laemmel07-POPL} in the context of applying `Scrap Your
Boilerplate' to XML programming. As we noted before in
\S\ref{S:adviceTypecase}, the use of type-class-based programming
involves substantial encoding, and hence, future type systems should
provide more direct means for expressing reachability contracts.

\mySection{Static analysis of traversal strategies}
\label{S:analysis}

We go beyond the limitations of established means of static typing by
proposing designated static analyses to curb the identified
programming errors. In this manner, we can address some problems more
thoroughly than with established means of static typing.  For
instance, we can infer contracts regarding fallibility and
reachability---as opposed to checking them previously. Also, the
encoding burden of the previous section would be eliminated by a
practical type system which includes the proposed analyses. Further,
we can address additional problems that were out of reach so far. In
particular, we can perform designated forms of termination analysis to
rule out divergent traversal.

The section is organized as a sequel of contributions---each of them
consisting of a piece of language-agnostic advice for improving strategic
programming and an associated static analysis to support the advice.

We use abstract interpretation and special-purpose type systems for
the specification and implementation of the analyses. We have included
a representative example of a soundness proof. In all cases, we have
modeled the analyses algorithmically in Haskell. All but some routine
parts of the analyses are included into the text; a cursory
understanding of the analyses does not require Haskell proficiency.


\newcommand{\adviceInferFallibility}{Perform fallibility analysis}
\mySubsection{\adviceInferFallibility}
\label{S:adviceInferFallibility}

\begin{advice} 
\label{A:inferFallibility}
Curb programming errors due to subtle control flow (see
\S\ref{S:cflow}) by statically analyzing strategy properties regarding
fallibility, i.e., success and failure behavior. Favorable properties
may be stated as contracts in programs which are verified by the
proposed analysis.
\end{advice}

Without loss of generality, we focus here on an analysis that
determines whether a strategy can be guaranteed to succeed (read as
`always succeeds' or `infallible'). Similar analyses are conceivable
for cases such as `always fails', `sometimes succeeds', and others .

We will first apply abstract interpretation to the problem, but come
to the conclusion that the precision of the analysis is insufficient
to yield any non-trivial results. We will then apply a special-purpose
type system; the latter approach provides sufficient precision. The
first approach nevertheless provides insight into success and failure
behavior, and the overall framework for abstract interpretation can
be later re-purposed for another analysis.


\mySubsubsection{An abstract interpretation-based approach}

We use the following lattice for the abstract domain for a simple
success and failure analysis.\footnote{We use the general framework of
  abstract interpretation by Cousot and
  Cousot~\cite{CousotC04,Cousot96}; we are specifically guided by
  Nielson and Nielson's style as used in their
  textbooks~\cite{NielsonNH05,NielsonN07}.}

\begin{center}
\includegraphics[height=3cm]{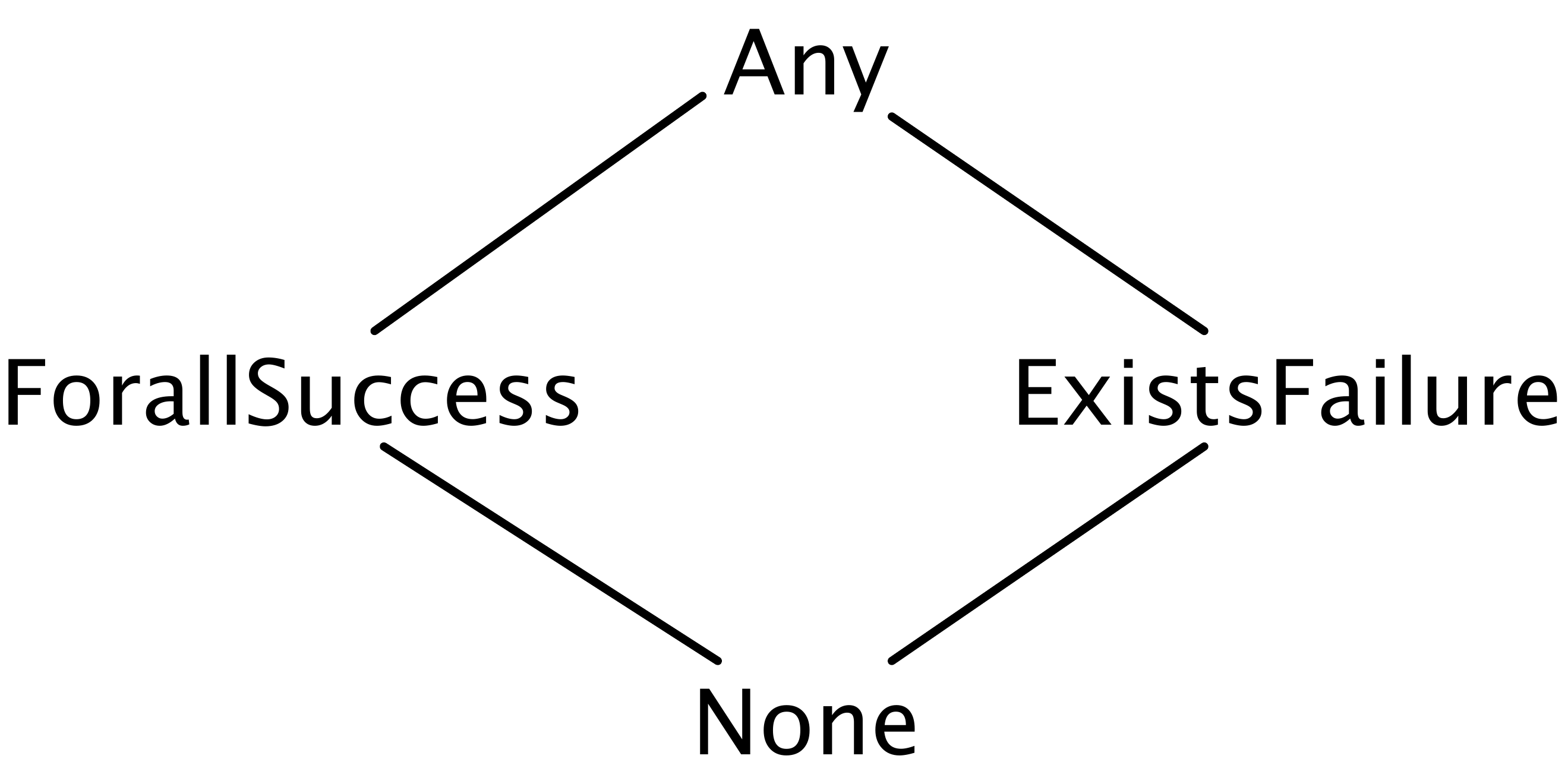}
\end{center}


\begin{figure}[t!]
\begin{boxedminipage}{\hsize}
\begin{lstlisting}[aboveskip=3pt]
-- General framework for abstract domains

class Eq x => POrd x
 where
  (<=) :: x -> x -> Bool
  (<) :: x -> x -> Bool
  x < y = not (x==y) && x <= y

class POrd x  => Bottom x    where bottom :: x
class POrd x  => Top x       where top :: x
class POrd x  => Lub x       where lub :: x -> x -> x


-- The abstract domain for success and failure analysis

data Sf = None | ForallSuccess | ExistsFailure | Any

instance POrd Sf
 where
  None <= _     = True
  _    <= Any   = True
  x    <= y     = x == y

instance Bottom Sf   where bottom = None
instance Top Sf      where top = Any

instance Lub Sf 
 where
  lub None x      = x
  lub x    None   = x
  lub Any  x      = Any
  lub x    Any    = Any
  lub x    y      = if x == y then x else Any
\end{lstlisting}
\end{boxedminipage}
\caption{\textbf{The abstract domain for success and failure behavior.}}
\label{F:SfDomain}
\figskip
\end{figure}


The bottom element \hs{None}\ represents the absence of any
information, and gives the starting point for fixed point
iteration. The two data values above \hs{None} represent the following
cases:
\begin{itemize}
\item There is no value where the strategy fails: \hs{ForallSuccess} 
\item There is a failure point for the strategy: \hs{ExistsFailure}.
\end{itemize} 
In the former case, we also speak of an \emph{infallible} strategy
according to \S\ref{S:properties}. Note that this is a \emph{partial
  correctness} analysis. Hence, in none of the cases is it implied
that the program terminates for all arguments. The `top' value,
\hs{Any}, represents the result of an analysis that concludes with
both of the above cases as being possible. Such a result tells us
nothing of value.

We refer to \autoref{F:SfDomain} for the Haskell model of the
abstract domain. We assume appropriate type classes for partial orders
(\hs{POrd}), least elements (\hs{Bottom}), greatest elements
(\hs{Top}), and least upper bounds (\hs{Lub}).


\begin{figure}[t!]
\begin{boxedminipage}{\hsize}
\begin{lstlisting}[aboveskip=3pt]
-- The actual analysis
analyse :: T Sf -> Sf
analyse Id            = ForallSuccess
analyse Fail          = ExistsFailure
analyse (Seq s s')    = analyse s `seq` analyse s'
analyse (Choice s s') = analyse s `choice` analyse s'
analyse (Var x)       = x
analyse (Rec f)       = fixEq (analyse . f)
analyse (All s)       = analyse s
analyse (One s)       = ExistsFailure

-- Equality-based fixed-point combinator
fixEq :: (Bottom x, Eq x) => (x -> x) -> x
fixEq f = iterate bottom
 where
  iterate x = let x' = f x
               in if (x==x') then x else iterate x'

-- Abstract interpretation of sequential composition
seq :: Sf -> Sf -> Sf
seq None          _             = None
seq ForallSuccess None          = None
seq ForallSuccess ForallSuccess = ForallSuccess
seq ForallSuccess _             = Any
seq ExistsFailure _             = ExistsFailure
seq Any           _             = Any

-- Abstract interpretation of left-biased choice
choice :: Sf -> Sf -> Sf
choice ForallSuccess _             = ForallSuccess
choice _             ForallSuccess = ForallSuccess
choice None          _             = None
choice _             None          = None
choice _             _             = Any
\end{lstlisting}
\end{boxedminipage}
\caption{\textbf{Abstract interpretation for analyzing the success and failure
  behavior of traversal programs.}}
\label{F:SfAnalysis}
\figskip
\end{figure}


The actual analysis is shown in full detail in
\autoref{F:SfAnalysis}. It re-interprets the abstract syntax of
\S\ref{S:interpreter} to perform abstract interpretation on the
abstract domain as opposed to regular interpretation. We discuss the
analysis case by case now. The base cases can be guaranteed to succeed
(\hs{Id}) and to fail (\hs{Fail}). The composition functions for the
compound strategy combinators can easily be verified to be monotone.

In the case of sequential composition, we infer success if both of the
operands succeed, while (definite) failure can be inferred only if the
first operand fails. The reason for this is that the failure point in
the domain of the second operand may not be in the range of the
first. Hence, we conclude with \hs{Any}\ in some cases. In the case of
choice, we infer success if either of the components succeeds, while
failure cannot be inferred at all. The reason for this is that two
failing operands may have different failure points. Hence, we have to
conclude again with the imprecise \hs{Any}\ value in some cases.

A reference to a \hs{Var}\ simply uses the information it contains,
and \hs{fixEq}\ is the standard computation of the least fixed point
within a lattice, by iteratively applying the function to the
\hs{bottom}\ element (in our analysis \hs{None}). We infer success for
an `all' traversal if the argument strategy is guaranteed to succeed;
likewise, the `all' traversal has a failure point if the argument
strategy has a failure point (because we could construct a term to
exercise the failure point on an immediate subterm position, assuming
a homogeneous as opposed to a many-sorted set of terms). We infer
\hs{ExistsFailure}\ for a `one' traversal because it has a failure
point for every constant term, regardless of the argument
strategy. \autoref{T:SfAnalysis} shows the results of the analysis.


\begin{figure}[t!]
{\footnotesize\noskip
\begin{center}
\begin{tabular}{|l|c|c|c|}
\hline 
 \   & \lstbox{s::ForallSuccess} & \lstbox{s::ExistsFailure} & \lstbox{s::Any} \\
\hline 
\lstbox{full_bu s} &\lstbox{None} &\lstbox{None} &\lstbox{None} \\
\lstbox{full_td s} & \lstbox{None} & \lstbox{ExistsFailure} & \lstbox{Any}\\
\lstbox{once_bu s} & \lstbox{ForallSuccess} & \lstbox{Any} & \lstbox{Any}\\
\lstbox{once_td s} & \lstbox{ForallSuccess} & \lstbox{Any} & \lstbox{Any}\\
\lstbox{stop_td s} & \lstbox{ForallSuccess} & \lstbox{None} & \lstbox{None}\\
\lstbox{innermost s} & \lstbox{ForallSuccess} & \lstbox{ForallSuccess} & \lstbox{ForallSuccess} \\
\hline
\end{tabular}
\end{center}}
\medskip
\caption{\textbf{Exercising success and failure analysis on traversal schemes.}}
\label{T:SfAnalysis}
\figskip
\end{figure}


The columns are labeled by the assumption for the success and failure
behavior of the argument strategy \hs{s}\ of the traversal
scheme. There is no column for \hs{None}\ since this value is only
used for fixed-point computation. There are a number of cells with the
\hs{None}\ value, which means that the analysis was not able to make
any progress during the fixed point computation. The analysis is
patently useless in such cases. There are a number of cells with the
\hs{Any}\ value, which means that the analysis concluded with an
imprecise result: we do not get to know anything of value about the
success and failure behavior in such cases.

All the cells with values \hs{ForallSuccess}\ and \hs{ExistsFailure}\
are as expected, but overall the analysis fails to recover behavior
in most cases. For instance, we know that a stop-top-down traversal is
guaranteed to succeed (see \S\ref{S:properties}), but the analysis
reports \hs{None}. 

One may want to improve the abstract interpretation-based approach so
that it computes more useful results. Here we note that the informal
arguments in support of fallibility and infallibility of traversal
schemes typically rely on induction over all possible terms. It is not
straightforward to adjust abstract interpretation in a way to account
for induction.

We abandon abstract interpretation for now. It turns out that a
type-system-based approach provides useful results rather easily
because it has fundamentally different characteristics of dealing with
recursion. In \S\ref{S:adviceProveReachability}, we will revisit
abstract interpretation and apply it successfully to a reachability
analysis for type-specific cases.


\mySubsubsection{A type system-based approach}
\label{S:sfTypes}

Let us use deduction based on a special-purpose type system to infer
when a strategy can be guaranteed always to yield a value \emph{if it
  terminates}, that is, it fails for no input. We use the type \true{}
for such a situation and \false{} for the lack thereof.


\begin{figure}[t!]

\begin{boxedminipage}{\hsize}

{\small

\ax{id^{\mbox{SF}}}{%
\cwtj{\Gamma}{\idT}{\true}
}

\ax{fail^{\mbox{SF}}}{%
\cwtj{\Gamma}{\failT}{\false}
}

\ir{sequ.1^{\mbox{SF}}}{%
\cwtj{\Gamma}{s_1}{\true}
\dnp
\cwtj{\Gamma}{s_2}{\true}
}{%
\cwtj{\Gamma}{\sequ{s_1}{s_2}}{\true}
}

\ir{sequ.2^{\mbox{SF}}}{%
\cwtj{\Gamma}{s_1}{\false}
\dnp
\cwtj{\Gamma}{s_2}{\tau}
}{%
\cwtj{\Gamma}{\sequ{s_1}{s_2}}{\false}
}

\ir{sequ.3^{\mbox{SF}}}{%
\cwtj{\Gamma}{s_1}{\tau}
\dnp
\cwtj{\Gamma}{s_2}{\false}
}{%
\cwtj{\Gamma}{\sequ{s_1}{s_2}}{\false}
}

\ir{choice.1^{\mbox{SF}}}{%
\cwtj{\Gamma}{s_1}{\false}
\dnp
\cwtj{\Gamma}{s_2}{\true}
}{%
\cwtj{\Gamma}{\choice{s_1}{s_2}}{\true}
}

\ir{choice.2^{\mbox{SF}}}{%
\cwtj{\Gamma}{s_1}{\false}
\dnp
\cwtj{\Gamma}{s_2}{\false}
}{%
\cwtj{\Gamma}{\choice{s_1}{s_2}}{\false}
}

\ir{choice.3^{\mbox{SF}}}{%
\cwtj{\Gamma}{s_1}{\true}
\dnp
\cwtj{\Gamma}{s_2}{\tau}
}{%
\cwtj{\Gamma}{\choice{s_1}{s_2}}{\true}
}

\ir{all^{\mbox{SF}}}{%
\cwtj{\Gamma}{s}{\tau}
}{%
\cwtj{\Gamma}{\all{s}}{\tau}
}

\ir{one^{\mbox{SF}}}{%
\cwtj{\Gamma}{s}{\tau}
}{%
\cwtj{\Gamma}{\one{s}}{\false}
}

\ir{rec^{\mbox{SF}}}{%
\cwtj{v:\tau,\Gamma}{s}{\tau}
}{%
\cwtj{\Gamma}{\rec{v}{s}}{\tau}
}

}

\end{boxedminipage}
\caption{\textbf{Typing rules for success and failure behavior}}
\label{F:SfTypes1}
\figskip
\end{figure}


The rules in \autoref{F:SfTypes1} describe a typing judgement such
that $\cwtj{\Gamma}{s}{\true}$ is intended to capture the judgement
that the strategy $s$ does not fail for any argument $t$, in the
context $\Gamma$. That is, there does not exist any $t$ such that
$\ioj{\apply{s}{t}}{\failure}$, according to the semantics in
\autoref{F:plus}--\autoref{F:minus}.


\begin{figure}[t!]
\begin{boxedminipage}{\hsize}
\begin{lstlisting}[aboveskip=3pt]
-- Type expressions
type Type = Bool -- Can we conclude that there is definitely no failure?

-- Type inference
typeOf :: T Type -> Maybe Type
typeOf Id             = Just True
typeOf Fail           = Just False
typeOf (Seq s s')     = liftM2 (&&) (typeOf s) (typeOf s')
typeOf (Choice s s')  = liftM2 (||) (typeOf s) (typeOf s')
typeOf (Var x)        = Just x
typeOf (Rec f)        = rec f True `mplus` rec f False
typeOf (All s)        = typeOf s
typeOf (One s)        = typeOf s >> Just False

-- Infer type of recursive closure
rec :: (Type -> T Type) -> Type -> Maybe Type
rec f t = typeOf (f t) >>= \t' ->
            if t==t' then Just t else Nothing
\end{lstlisting}
\end{boxedminipage}
\caption{\textbf{Type inference for success and failure behavior}}
\label{F:SfTypes2}
\figskip
\end{figure}


\autoref{F:SfTypes2} rephrases \autoref{F:SfTypes1} in a directly
algorithmic manner in Haskell---also providing type inference. The
context $\Gamma$, which carries fallibility information about free
variables in a term, is needed so that the analysis can deal with
recursively-defined strategies.

The property of infallibility is undecidable, and hence, the
type system will not identify all strategies of type \true, but it is
guaranteed to be sound, in that no strategy is mis-identified as being
infallible by the type system when it is not.

When we compare this approach to the abstract interpretation-based
approach, then \false{} should be compared with \hs{Any}\ as
opposed to \hs{ExistsFailure}. That is, \true{} represents
guarantee of success, while \false{} represents lack of such a
guarantee, as opposed to existence of a failure point. There is no
counterpart for \hs{ExistsFailure}\ in the type system. There is
certainly no counterpart for \hs{None}\ either, because this value is an
artifact of fixed point iteration, which is not present in the type
system.

With this comparison in mind, the deduction rules of
\autoref{F:SfTypes1} (and the equations of \autoref{F:SfTypes2}) are
very similar to the equations of \autoref{F:SfAnalysis}. For instance,
the rules for base cases $\idT^{\mbox{SF}}$ and $\failT^{\mbox{SF}}$ state
that the identity, \idT, is infallible, but that the primitive
failure, \failT, is not. For a sequence to be infallible, both
components need to be infallible ($sequ.1^{\mbox{SF}}$ to
$sequ.3^{\mbox{SF}}$). If either component of a choice is infallible,
the choice is too ($choice.1^{\mbox{SF}}$, $choice.3^{\mbox{SF}}$). A
choice between two potentially fallible programs might well be
infallible, but this analysis can only conclude that this is not
guaranteed, and it is here that imprecision comes into the
analysis. The type of an `all' traversal coincides with the type of
argument strategy ($all^{\mbox{SF}}$). There is no guarantee of
success for a `one' traversal ($one^{\mbox{SF}}$).

Finally, in dealing with the recursive case it is necessary to
introduce a type context, $\Gamma$, containing typing assertions on
variables. To conclude that a recursive definition $\rec{v}{s}$ is
infallible, it is sufficient to show that the body of the recursion,
$s$, is infallible assuming that the recursive call, $v$, is too.


\begin{figure}[t!]
{\footnotesize\noskip
\begin{center}
\begin{tabular}{|l|c|c|}
\hline 
 \   & \mathbox{s::\false} & \mathbox{s::\true} \\
\hline
\lstbox{full_bu s} &\mathbox{\false} &\mathbox{\true} \\
\lstbox{full_td s} & \mathbox{\false} & \mathbox{\true} \\
\lstbox{once_bu s} & \mathbox{\false} & \mathbox{\true} \\
\lstbox{once_td s} & \mathbox{\false} & \mathbox{\true} \\
\lstbox{stop_td s} & \mathbox{\true} & \mathbox{\true} \\
\lstbox{innermost s} & \mathbox{\true} & \mathbox{\true} \\
\hline
\end{tabular}
\end{center}}
\medskip
\caption{\textbf{Exercising success and failure types on traversal schemes.}}
\label{T:SfTypes}
\figskip
\end{figure}


\autoref{T:SfTypes} presents the results of using the type system for
some common traversals. Again, the columns label the assumption for
the success and failure behavior of the argument strategy \hs{s}\ of
the traversal scheme. (We use the context parameter of the type
system, or, in fact, the \hs{Var}\ form of strategy terms, to capture
and propagate such assumptions; see the paper's online code
distribution.)  When compared to \autoref{T:SfAnalysis}, guarantee of
success is inferred for several more cases. For instance, such a
guarantee is inferred for the schemes of full top-down and bottom-up
traversal, subject to the guarantee for the argument. Also, the scheme
for stop-top-down traversal is found to universally succeed, no matter
what the argument strategy. The abstract interpretation-based approach
could not make a useful prediction for these cases.


\mySubsubsection{Simple dead-code detection}

As an aside, there is actually a trivial means to improve the
usefulness of the type system. That is, we can easily exclude certain
strategies that involve dead code in the sense of \S\ref{S:dead-code}.
Specifically, we could remove the following rule:

{\small

\ir{choice.3^{\mbox{SF}}}{%
\cwtj{\Gamma}{s_1}{\true}
\dnp
\cwtj{\Gamma}{s_2}{\tau}
}{%
\cwtj{\Gamma}{\choice{s_1}{s_2}}{\true}
}

}

In this way, we classify a choice construct $\choice{s_1}{s_2}$ with
an infallible left operand as ill-typed---the point being that the
composition is equivalent to $s_{1}$ with $s_{2}$ being dead code.


\mySubsubsection{Soundness of the type system}
\label{S:soundness}

We prove soundness of the type system in \autoref{F:SfTypes1} relative
to the established, natural semantics in
\autoref{F:plus}--\autoref{F:minus}.

\begin{thm}
For all strategic programs $s$ if \cwtj{}{s}{\true} then for no term $t$  \ioj{\apply{s}{t}}{\failure}.
\end{thm}

\vspaceSection

\paragraph*{Proof}

We use a proof by contradiction.  We suppose that there is some
program $s$ such that \cwtj{}{s}{\true} and that there is an argument
$t$ so that \ioj{\apply{s}{t}}{\failure}, and we choose $s$ and $t$ so
that the depth of the derivation of \ioj{\apply{s}{t}}{\failure} is
minimal; from this we derive a contradiction. We work by cases over
$s$.

\begin{description}

\item[Identity] If $s$ is \idT{} then there is no evaluation rule
  deriving \ioj{\apply{\idT}{t}}{\failure} for any $t$,
  contradicting the hypothesis.

\smallskip

\item[Failure] If $s$ is \failT{} then there is no typing rule
  deriving \cwtj{}{\failT}{\true}, contradicting the hypothesis.

\smallskip

\item[Sequence] If $s$ is $\sequ{s_{1}}{s_{2}}$ then the only way that
  \cwtj{}{\sequ{s_{1}}{s_{2}}}{\true} can be derived is for the typing rule
  $sequ.1^{\mbox{SF}}$ to be applied to derivations of
  \cwtj{}{s_{1}}{\true} and \cwtj{}{s_{2}}{\true}.

  Now, by hypothesis we also have a term $t$ so that
  \ioj{\apply{\sequ{s_{1}}{s_{2}}}{t}}{\failure}: examining the evaluation
  rules we see that this can only be deduced from
  \ioj{\apply{s_{1}}{t}}{\failure} by rule $seq^-.1$ or from
  \ioj{\apply{s_{2}}{t}}{\failure} by rule $seq^-.2$.

  We choose $i$ such that \ioj{\apply{s_{i}}{t}}{\failure}; the
  corresponding derivation is shorter than
  \ioj{\apply{s}{t}}{\failure}, a contradiction to the minimality of
  the derivation for $s$.

\smallskip

\item[Choice] If $s$ is $\choice{s_{1}}{s_{2}}$ then there are two
  ways that \cwtj{}{\choice{s_{1}}{s_{2}}}{\true} can be derived:
  using $choice.3^{\mbox{SF}}$ from a derivation of
  \cwtj{}{s_{1}}{\true} or using $choice.1^{\mbox{SF}}$ from a
  derivation of \cwtj{}{s_{2}}{\true}.

  Now, by hypothesis we also have a term $t$ so that
  \ioj{\apply{\choice{s_{1}}{s_{2}}}{t}}{\failure}: examining the
  evaluation rules we see that this can only be deduced from
  \ioj{\apply{s_{1}}{t}}{\failure} and
  \ioj{\apply{s_{2}}{t}}{\failure} by rule $choice^-$.

  We choose $s_{i}$ to be the case where
  $\cwtj{}{s_{i}}{\true}$. Whichever we choose, the derivation of
  \ioj{\apply{s_{i}}{t}}{\failure} is shorter than
  $\ioj{\apply{s}{t}}{\failure}$, a contradiction to the minimality of
  the derivation for $s$.

\smallskip

\item[All] If $s$ is $\all{s'}$ then $\cwtj{}{\all{s'}}{\true}$ is
  derived from $\cwtj{}{s'}{\true}$. From the negative rules for
  evaluation we conclude that $t$ is of the form $c(t_{1}, \ldots,
  t_{n})$ and for some $i$ we have
  $\ioj{\apply{s'}{t_{i}}}{\failure}$, and the derivation of this will
  be shorter than that of $\ioj{\apply{s}{t}}{\failure}$, in
  contradiction to the hypothesis.

\smallskip

\item[One] If $s$ is $\one{s'}$ then $\cwtj{}{\one{s'}}{\true}$ cannot
  be derived, directly contradicting the hypothesis.

\smallskip

\item[Recursion] Finally we look at the case that $s$ is of the form
  $\rec{v}{s'}$. We have a derivation ($d_{1}$, say) of
  $\cwtj{}{\rec{v}{s'}}{\true}$, and this is constructed by applying
  rule $rec^{\mbox{SF}}$ to a derivation $d_{2}$ of
  $\cwtj{v:\true}{s'}{\true}$.

  We also have the argument $t$ so that
  $\ioj{\apply{\rec{v}{s'}}{t}}{\failure}$. This in turn is derived
  from a derivation
  \ioj{\apply{s'[v\mapsto\rec{v}{s'}]}{t}}{\failure}, shorter than the
  former. So, $s'[v\mapsto\rec{v}{s'}]$ will be our counterexample to
  the minimality of $\rec{v}{s'}$, so long as we can derive
  $\cwtj{}{s'[v\mapsto\rec{v}{s'}]}{\true}$.

  We construct a derivation of this from $d_{2}$, replacing each
  occurrence in $d_{2}$ of the variable rule applied to $v:\true$ by a
  copy of the derivation $d_{1}$, which establishes that the value
  substituted for $v$, $\rec{v}{s'}$, has the type $\true$, thus
  giving a derivation of $\cwtj{}{s'[v\mapsto\rec{v}{s'}]}{\true}$, as
  required to prove the contradiction.
\end{description}


\newcommand{\adviceProveReachability}{Perform reachability analysis}
\mySubsection{\adviceProveReachability}
\label{S:adviceProveReachability}

\begin{advice} 
\label{A:proveReachability}
Curb programming errors due to type-specific cases not exercised
during traversal (see \S\ref{S:dead-code}) by statically analyzing
reachability of the cases within strategies that are applied to a term
of a statically known type. Such dead code detection does not require
programmer-provided reachability contracts; instead it is a general
analysis of strategy applications.
\end{advice}


\begin{figure}[t!]
{\footnotesize\noskip
\begin{center}
\begin{tabular}{|l|l|c|c|}
\hline 
& Strategy & Root type & Reachable type-specific cases \\
\hline 
1. & \lstbox{Id} & \lstbox{Company} & \ensuremath{\emptyset} \\
2. & \lstbox{incSalary} & \lstbox{Salary} & \ensuremath{\{\lstbox{incSalary}\}} \\
3. & \lstbox{try incSalary} & \lstbox{Employee} & \ensuremath{\emptyset} \\
4. & \lstbox{All\ (try incSalary)} & \lstbox{Employee} & \ensuremath{\{\lstbox{incSalary}\}} \\
5. & \lstbox{All\ (try\ incSalary)} & \lstbox{Department} & \ensuremath{\emptyset} \\
6. & \lstbox{once_bu\ (try\ incSalary)} & \lstbox{Department} & \ensuremath{\{\lstbox{incSalary}\}} \\
\hline
\end{tabular}
\end{center}}
\medskip
\caption{\textbf{Exercising the reachability analysis for companies.}}
\label{T:Reach}
\figskip
\end{figure}


For instance, using again the introductory company example, we would
like to obtain the kind of information in \autoref{T:Reach} by a
reachability analysis. In this example, we assume one type-specific
case, \lstbox{incSalary}, which is used in the traversal program
subject to the analysis. The case increases salaries and we assume
that it is applicable to salary terms only.

Let us motivate some of the expected results in detail. When applying
the strategy \hs{Id}\ (see the first line of  the figure), which clearly
does not involve any type-specific case, we obtain the empty set of
reachable cases. When applying the type-specific \hs{incSalary}\ to a
salary term (second line), then the case is indeed applied; hence the
result is $\{\mbox{\hs{incSalary}}\}$. We cannot usefully apply
\hs{incSalary}\ to an employee (third line); hence, we obtain the
empty set of reachable cases. We may though apply  \hs{All (try
  incSalary)}\ to an employee (fourth line) because a salary may be
encountered as an immediate subterm of an employee. The last
two lines in the table illustrate the reachability behavior of a
traversal scheme, in comparison to a one-layer traversal.


\begin{figure}[t!]
\begin{boxedminipage}{\hsize}
\begin{lstlisting}[aboveskip=3pt]
-- Representation of signatures
type Sort      = String
type Constr    = String
type Symbol    = (Constr,[Sort],Sort)
data Signature = Signature { sorts   :: Set Sort
                           , symbols :: Set Symbol
                           }

-- Additional observer functions
argSortsOfSort :: Signature -> Sort -> Set Sort
...
\end{lstlisting}
\end{boxedminipage}
\caption{\textbf{An abstract data type for signatures.}}
\label{F:Types}
\figskip
\end{figure}


The abstract interpretation relies on many-sorted signatures for the
terms to be traversed, as shown in \autoref{F:Types}; we omit the
straightforward definition of various constructors and observers. In
the code for the abstract interpretation for reachability, we will
only use the observer \lstbox{sorts} of type \lstbox{Signature -> Set
  Sort}, to retrieve all possible sorts of a signature, and the
observer \lstbox{argSortsOfSort}, to retrieve all sorts of immediate
subterms for all possible terms of a given sort.

For simplicity's sake, we formally represent type-specific cases
simply by their name. Reachability analysis returns sets of such cases
(say, names or strings). Hence we define:

\begin{lstlisting}
type Case  = String
type Cases = Set Case
\end{lstlisting}

For each case, we need to capture its `sort'. Only when a
type-specific case is applied to a term of the designated sort, then
the case should be counted as being exercised. More generally:
\emph{The abstract interpretation computes what
  type-specific cases are exercised by a given strategy when faced
  with terms of different sorts}. To this end, we use the following
abstract domain:

\begin{lstlisting}
type Abs = Map Sort Cases
\end{lstlisting}

Here, \hs{Map}\ is a Haskell type for finite maps: sorts are
associated with type-specific cases. The analysis associates each
strategy with such a map---as evident from the following function
signature:

\begin{lstlisting}
analyse :: Signature -> T Abs -> Abs
\end{lstlisting}

That is, for each \hs{Sort}\ in the given signature the analysis is
supposed to return a set of (named) type-specific \hs{cases}\ which
\emph{may} be executed by the traversal if the traversal is applied to
a term of the sort. For instance:

\begin{lstlisting}
> let incSalary = fromList [("Salary",Set.fromList ["incSalary"])]
> analysis companySignature (All (All (Var incSalary)))
[("Unit",fromList ["incSalary"]),("Manager",fromList ["incSalary"])]
\end{lstlisting}

The first input line shows the assembly of a type-specific case which
is represented here as a trivial map of type \hs{Abs}. The second
input line starts a reachability analysis. The printed map for the
result of the analysis states that \hs{incSalary} can be reached from
both \hs{Unit}\ and \hs{Manager}. Indeed, salary components occur
exactly two constructor levels below \hs{Unit}\ and \hs{Manager}.

The analysis is safe in that it is guaranteed to return all cases
which are executed on some input; it is however an over-approximation,
and so no guarantee is provided that all returned cases are actually
executed.


\begin{figure}[t!]
\begin{boxedminipage}{\hsize}
\begin{lstlisting}[aboveskip=3pt]
analyse :: Signature -> T Abs -> Abs
analyse sig = analyse' 
 where
  analyse' :: T Abs -> Abs
  analyse' Id            = bottom
  analyse' Fail          = bottom
  analyse' (Seq s s')    = analyse' s `lub` analyse' s'
  analyse' (Choice s s') = analyse' s `lub` analyse' s'
  analyse' (Var x)       = x
  analyse' (Rec f)       = fixEq (analyse' . f)
  analyse' (All s)       = transform sig $ analyse' s
  analyse' (One s)       = transform sig $ analyse' s

transform :: Signature -> Abs -> Abs
transform sig abs
 = Map.fromList 
 $ map perSort 
 $ Set.toList 
 $ sorts sig
 where
  perSort :: Sort -> (Sort, Cases)
  perSort so = (so,cases)
   where
    cases = lubs
          $ map perArgSort
          $ Set.toList argSorts
     where
      argSorts = argSortsOfSort sig so
      perArgSort = flip Map.lookup abs
\end{lstlisting} 
\end{boxedminipage}
\caption{\textbf{Abstract interpretation for analyzing the reachability of
type-specific cases relative to a given signature.}}
\label{F:Reach}
\figskip
\end{figure}


The analysis proceeds by induction over the structure of strategies,
and is parametrized by the \hs{Signature}\ over which the strategy is
evaluated. The analysis crucially relies on the algebraic status of
finite maps to define partial orders with general least upper bounds
subject to the co-domain of the maps being a lattice itself. (We use
the set of all subsets of type-specific cases as the co-domain.) The
bottom value of this partial order is the map that maps all values of
the domain to the bottom value of the co-domain. Here we note that
such maps are an established tool in static program analysis. For
instance, maps may be used in the analysis of imperative programs for
\emph{property states} as opposed to concrete states for program
variables as in \cite{NielsonNH05,NielsonN07}.

The central part of the analysis is the treatment of the one-layer
traversals \hs{All s}\ and \hs{One s}. The reachable cases are
determined separately for each possible sort, and these per-sort
results are finally combined in a map. For each given sort \hs{so},
the recursive call of the analysis, \hs{analyse' s}, is exercised
for all possible sorts of immediate subterms of terms of sort
\hs{so}. Fixed point iteration will eventually reach all reachable
sorts in this manner.


The analysis is conservative in so far that it distinguishes neither
\hs{Seq}\ from \hs{Choice}\ nor \hs{All}\ from \hs{One}\ in any
manner. Also, the analysis assumes all constructors to be universally
feasible, which is generally not the case due to type-specific cases
and their recursive application---think of the patterns in rewrite
rules. As a result, certain reachability-related programming errors
will go unnoticed. Consider the following example:

\begin{lstlisting}
stop_td (Choice leanDepartment (try incSalary)) myCompany
\end{lstlisting}

For the sake of a concrete intuition, we assume that
\hs{leanDepartment}\ will pension off all eligible employees, if any.
Here we assume that \hs{leanDepartment}\ applies to terms of sort
\hs{Department} and it succeeds for all such terms. As a result of
\hs{leanDepartment}'s success at the department level of company
terms, the stop-top-down traversal will never actually hit employees
or salaries that are only to be found below departments.

This illustration clearly suggests that a success and failure analysis
should be incorporated into the reachability analysis for precision's
sake. To this end, it would be beneficial to know whether a
type-specific case universally succeeds for all terms of the sort in
question. Further, the analysis should treat sequence differently from
choice, and an `all' traversal differently from a `one' traversal. We omit
such elaborations here.


\newcommand{\adviceTermination}{Perform termination analysis}
\mySubsection{\adviceTermination}
\label{S:adviceTermination}

\begin{advice} 
\label{A:termination}
Curb various kinds of programming errors that may manifest themselves
as divergent traversals. These includes wrong decisions regarding the
traversal scheme and misunderstood properties of rewrite rules. To this
end, perform a static termination analysis that leverages appropriate
measures in proving termination of recursive traversals.
\end{advice}

That is, we seek an analysis that determines, conservatively, whether
a given recursive strategy is guaranteed to converge. For instance,
the intended analysis should infer the following convergence
properties. A full bottom-up traversal converges regardless of the
argument strategy, as long as the argument strategy itself converges
universally. A full top-down traversal converges as long as the
argument strategy converges universally and does not increase some
suitable measure such as the depth of the term.


\begin{figure}[t!]
\begin{boxedminipage}{\hsize}
\begin{center}
\includegraphics[width=9cm]{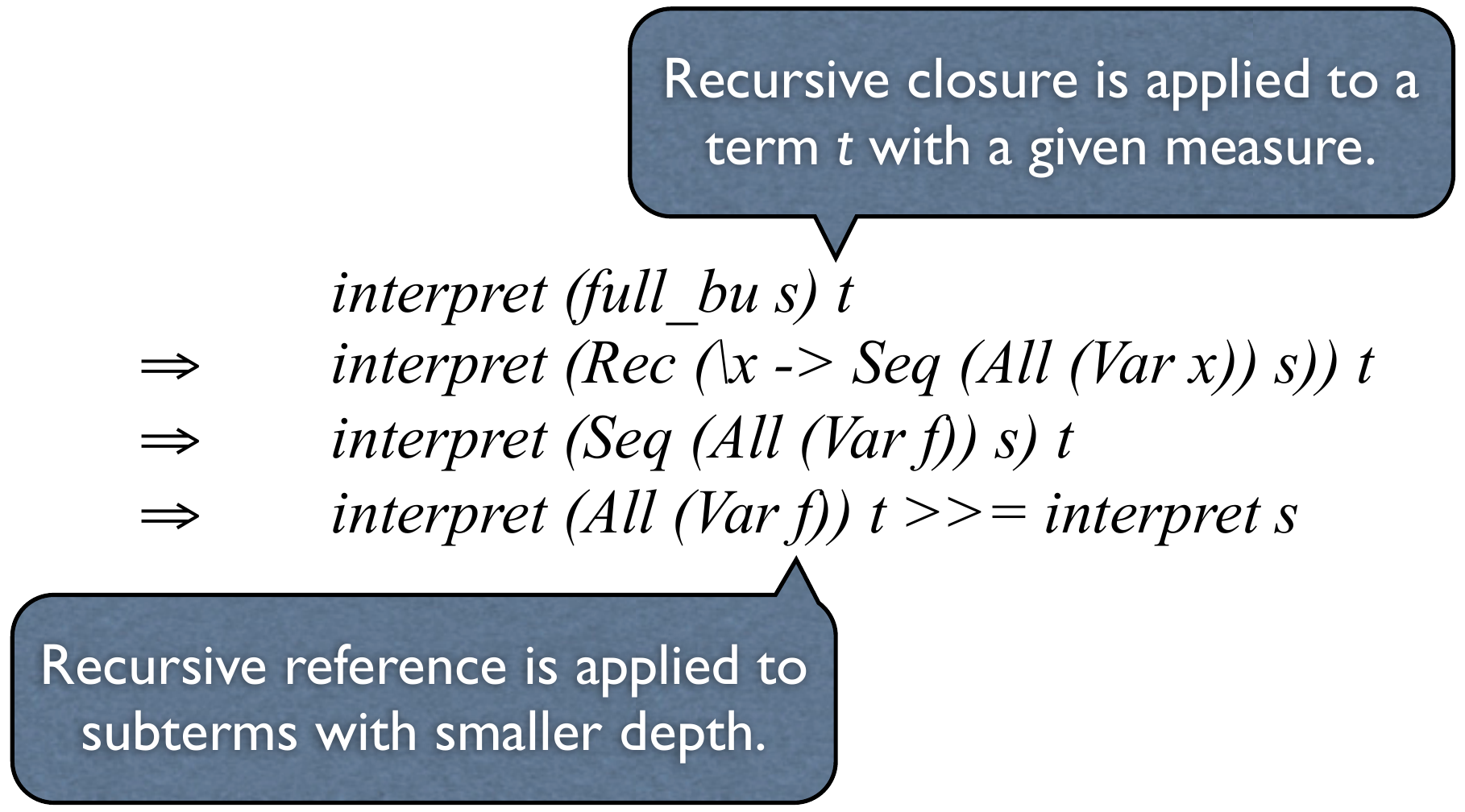}
\end{center}
\end{boxedminipage}
\caption{\textbf{Illustration of termination checking.}}
\label{F:Rel1}
\figskip
\end{figure}


Based on experiences with modeling strategies in
Isabelle/HOL~\cite{KaiserL09}, our analysis for termination checking
essentially leverages an induction principle; see \autoref{F:Rel1} for
an illustration. That is, the analysis is meant to verify
that a measure of the term (such as the depth of the term), as seen by
the recursive closure as a whole, is \emph{decreased} throughout the
body of the recursive closure until the recursive reference is
invoked. The figure shows a few steps of interpretation when applying
a full bottom-up traversal to a term. The recursive reference is
eventually applied to immediate subterms of the original term. Hence,
the mere depth measure of terms is sufficient here for the induction
principle.

We will first develop a basic termination analysis that leverages the
depth measure. However, many traversals in strategic programming
cannot be proven to converge just on the grounds of depth. For
instance, a program transformation for macro expansion could be
reasonably implemented with a full top-down traversal, but such a
traversal clearly increases term depth. Accordingly, we will
generalize the analysis to deal with measures other than depth.


\mySubsubsection{Measure relations on terms and strategies}

The key concept of the termination analysis is a certain style of
manipulating measures symbolically. We will explain this concept here
for the depth measure for ease of understanding, but the style
generalizes easily for other measures.

Based on the intuition of \autoref{F:Rel1}, the analysis must track the
depth measure of terms throughout the symbolic execution of a strategy
so that the depth can be tested at the point of recursive
reference. That is, \emph{the depth must be shown to be smaller at the
  point of recursive reference than the point of entering the
  recursive closure.} In this manner, we establish that the depth
measure is smaller for each recursive unfolding, which implies
convergence.


\begin{figure}[t!]
\begin{boxedminipage}{\hsize}
\begin{lstlisting}[aboveskip=3pt]
-- Relation on measures
data Rel = Leq | Less | Any

-- Partial order and LUB for Rel
instance POrd Rel
 where
  _    <= Any   = True
  Less <= Leq   = True  
  r    <= r'    = r == r'

instance Lub Rel
 where
  lub Less Leq  = Leq
  lub Leq  Less = Leq
  lub r    r'   = if r == r' then r else Any

-- Rel arithmetic
plus :: Rel -> Rel -> Rel
plus Less Less = Less
plus Less Leq  = Less
plus Leq  Less = Less
plus Leq  Leq  = Leq
plus _    _    = Any

decrease :: Rel -> Rel
decrease Leq  = Less
decrease Less = Less
decrease Any  = Any

increase :: Rel -> Rel
increase Less = Leq
increase Leq  = Any
increase Any  = Any
\end{lstlisting}
\end{boxedminipage}
\caption{\textbf{Relation on measures such as depth of terms.}}
\label{F:Rel2}
\figskip
\end{figure}


Obviously, the analysis cannot use actual depth, but it needs to use
an abstract domain. We use a finite domain \hs{Rel} whose values
describe the relation between the depths of two terms: i) the term in
a given position of the body of a recursive closure; ii) the term at
the entrance of the recursive closure. The idea is that the relation
is initialized to `$\leq$' (or `$=$' if the abstract domain provided
this option) as we enter the recursive closure, and it is accordingly
updated as we symbolically interpret the body of the
closure. Ultimately, we are interested in the relation at the
point of recursive reference. These are the values of \hs{Rel}; see
\autoref{F:Rel2} for a full specification:

\begin{lstlisting}
-- Relation on measures
data Rel = Leq | Less | Any
\end{lstlisting}

One can think of the values as follows. The value \hs{Leq}\ models that
the depth of the term was not increased during the execution of the
body of the recursive closure so far. The value \hs{Less}\ models that
the depth of the term was strictly decreased instead. This is the
relation that must hold at the point of the recursive reference. The
value \hs{Any}\ models that we do not know for certain whether the
depth was preserved, increased, or decreased.

So far we have emphasized a depth relation for \emph{terms}. However,
the analysis also relies on depth relations for
\emph{strategies}. That is, we can use the same domain \hs{Rel}\ to
describe \emph{the effect of a strategy on the depth}. In this case,
one can think of the values as follows. The value \hs{Leq}\ models
that the strategy does not increase the depth of the term. The value
\hs{Less}\ models that the strategy strictly decreases the depth of
term. The value \hs{Any}\ models that we do not know for certain
whether the strategy preserves, increases, or decreases depth.

For clarity, we use these type synonyms:

\begin{lstlisting}
type TRel = Rel  -- Term property
type SRel = Rel  -- Strategy property
\end{lstlisting}

We set up the type of the analysis as follows:

\begin{lstlisting}
type Abs = TRel -> Maybe TRel
analyse :: T SRel -> Abs
\end{lstlisting}

In fact, we would like to use variables of the \hs{T}\ type again to
capture effects for \emph{arguments} of traversal schemes. Hence, we
should distinguish recursive references from other references; we use
an extra Boolean to this end; \hs{True}\ encodes recursive references.
Thus:

\begin{lstlisting}
analyse :: T (SRel,Bool) -> Abs
\end{lstlisting}

At the top level, we begin analyzing a strategy expression (presumably
a recursive closure) by assuming a \hs{TRel}\ value of \hs{Leq}. Also,
we effectively provide type inference in that we compute an \hs{SRel}
value for the given strategy expression. Thus:

\begin{lstlisting}
typeOf :: T (SRel,Bool) -> Maybe SRel
typeOf s = analyse s Leq
\end{lstlisting}


\begin{figure}[t!]
\begin{boxedminipage}{\hsize}
\begin{lstlisting}[aboveskip=3pt]
analyse :: T (SRel, Bool) -> Abs
analyse Id = Just
analyse Fail = const (Just Less)
analyse (Seq s s') = maybe Nothing (analyse s') . analyse s

analyse (Choice s s')
 = \r ->
     case (analyse s r, analyse s' r) of
       (Just r1, Just r2) -> Just (lub r1 r2)
       _                  -> Nothing

analyse (Var (r,b))
 = \r' ->
     if not b || r' < Leq
       then Just (plus r' r)
       else Nothing

analyse (Rec f)
 = \r -> maybe Nothing (Just . plus r) (typeOfClosure r)
 where
  typeOfClosure r = if null attempts
                      then Nothing 
                      else Just (head attempts)
   where
    attempts = catMaybes (map wtClosure' [Less,Leq,Any])
    wtClosure r = maybe False (<=r) (analyse (f (r,True)) Leq)
    wtClosure' r = if wtClosure r then Just r else Nothing

analyse (All s) = transform (analyse s)
analyse (One s) = transform (analyse s)

transform :: Abs -> Abs
transform f r = maybe Nothing (Just . increase) (f (decrease r))
\end{lstlisting}
\end{boxedminipage}
\caption{\textbf{A static analysis for termination checking relative to the
  depth measure for terms.}}
\label{F:LeqTypes1}
\figskip
\end{figure}


\mySubsubsection{Termination analysis with the depth measure}

The analysis is defined in \autoref{F:LeqTypes1}. The analysis can be
viewed as an algorithmically applicable type system which tracks
effects on measures as types. Just in the same way as the standard
semantics threads a term through evaluation, this analysis threads a
term property of type \hs{TRel}\ through symbolic evaluation. The case
for \hs{Id}\ preserves the property (because \hs{Id}\ preserves the
depth, in fact, the term). The case for \hs{Fail}\ decreases depth in
a vacuous sense. The case for \hs{Seq}\ sequentially composes the
effects for the operand strategies. The case for \hs{Choice}\ takes
the least upper bound of the effects for the operand strategies. For
instance, if one operand possibly increases depth, then the composed
strategy is stipulated to potentially increase depth.

The interesting cases are those for variables (including the case of
recursive references), recursive closures and one-layer traversal. The
case for \hs{Var}\ checks whether we face a recursive reference (i.e.,
\hs{b == True}) because in this case we must insist on the current
\hs{TRel}\ value to be \hs{Less}. If this precondition holds, then the
strategy property for the variable is combined with the current term
property using the \hs{plus} operation (say, addition) for type
\hs{Rel}.

The case for \hs{Rec}\ essentially resets the \hs{TRel}\ value to
\hs{Leq}; see \hs{analyse ... Leq}, and attempts the analysis of the
body for all possible assumptions about the effect of the recursive
references; see \hs{[Less,Leq,Any]}. If the analysis returns with
any computed effect, then this result is required to be less or equal
to the one assumed for the recursive reference; see `\hs{<=}'. The
ordering on the attempts implies that the least constraining type is
inferred (i.e., the largest value of \hs{SRel}).

Finally, the cases for \hs{All}\ and \hs{One}\ are handled identically
as follows. The term property is temporarily decreased as the argument
strategy is symbolically evaluated, and the resulting term property is
again increased on the way out. This models the fact that argument
strategies of `all' and `one' are only applied to subterms. It is
important to understand that the operations increase and decrease are
highly constrained. In particular, once the \hs{TRel} value has
reached \hs{Any}, decrease does not get us back onto a
termination-proven path.


\begin{figure}[t!]
{\footnotesize\noskip
\begin{center}
\begin{tabular}{|l|c|c|c|}
\hline 
 \   & \lstbox{s::Any} & \lstbox{s::Leq} & \lstbox{s::Less} \\
\hline 
\lstbox{full_bu s}   & \lstbox{Just Any} & \lstbox{Just Leq} & \lstbox{Just Less}\\
\lstbox{full_td s}   & \lstbox{Nothing}  & \lstbox{Just Leq} & \lstbox{Just Leq}\\
\lstbox{stop_td s}   & \lstbox{Just Any} & \lstbox{Just Leq} & \lstbox{Just Leq}\\
\lstbox{once_bu s}   & \lstbox{Just Any} & \lstbox{Just Leq} & \lstbox{Just Leq}\\
\lstbox{repeat s}    & \lstbox{Nothing}  & \lstbox{Nothing}  & \lstbox{Just Leq}\\
\lstbox{innermost s} & \lstbox{Nothing}  & \lstbox{Nothing}  & \lstbox{Nothing}\\
\hline
\end{tabular}
\end{center}}
\medskip
\caption{\textbf{Exercising the termination analysis on traversal schemes.}}
\label{T:LeqTypes1}
\figskip
\end{figure}


The analysis is able to infer termination types for a range of
interesting traversal scenarios; see \autoref{T:LeqTypes1}. The table
clarifies that a full bottom-up traversal makes no assumption about
the measure effect of the argument strategy, while a full top-down
traversal does not get assigned a termination type for an
unconstrained argument; see the occurrence of \hs{Nothing}.

The depth measure is practically useless for typical applications of
\hs{repeat}, but we can observe nevertheless that an application of
\hs{repeat}\ will only terminate, if its argument is `strictly
decreasing'. This property is useful once we take into account other
measures.

At this point, we are not yet able to find any terminating use case
for \hs{innermost}. This is not surprising because \hs{innermost}\
composes \hs{repeat}\ and \hs{once_bu}, where the former requires a
strictly decreasing strategy (i.e., \hs{Less}), and the latter can
only be type-checked with \hs{Leq}\ as the termination type. Compound
measures, as discussed below, come to the rescue.


\mySubsubsection{Termination analysis with compound measures}

The static analysis can be elaborated to deal with measures other than
depth. In this paper, we demonstrate a measure of \emph{the number of
  occurrences of a specific constructor} in a term, i.e., the
constructor count. This sort measure is applicable to transformation
scenarios where a specific constructor is systematically eliminated,
e.g., in the sense of macro expansion.

This approach could also be generalized to deal with a measure for the
\emph{number of matches for a specific pattern} in a term, such as the
LHS of a rewrite rule. This elaboration is not discussed any further
though.

The general idea is to associate strategy expressions and argument
strategies of schemes in a traversal program with a suitable
measure. In this paper, we require that the programmer provides the
measure, but ultimately such measures may be inferred. We need a new
type, \hs{Measure}, to represent compound measure in a list-like
structure. Here, we assume that the depth measure always appears in
the last (least significant) position.

\begin{lstlisting}
data Measure = Depth | Count Constr Measure
type Constr = String
\end{lstlisting}

The type of measure transformation and the signature of the program
analysis must be changed such that we use non-empty lists of
values of type \hs{TRel}\ as opposed to singletons before, thereby
accounting for compound measures. Thus:

\begin{lstlisting}
type Abs = [TRel] -> Maybe [TRel]
analysis :: T ([SRel],Bool) -> Abs
\end{lstlisting}

The initial list of type \hs{[TRel]} is trivially obtained from 
(the length of) the measure by a function like this:

\begin{lstlisting}
leqs Depth = [Leq]
leqs (Count _ m) = Leq : leqs m
\end{lstlisting}

Thus, the analysis computes the effects of the strategy while assuming
that the declared measure holds for the argument. At this level of
development, the analysis is oblivious to the actual constructor names
because \hs{T}\ does not involve any expressiveness for dealing with
specific constructors. Measure claims are to be verified for given
rewrite rules that serve as arguments, but this part is omitted here.


\begin{figure}[t!]
{\footnotesize\noskip
\begin{center}
\begin{tabular}{|l|c|}
\hline 
 \   & \lstbox{s::[Less,Any]} \\
\hline 
\lstbox{full_td s}   & \lstbox{Just [Less,Any]}\\
\lstbox{once_bu s}   & \lstbox{Just [Less,Any]}\\
\lstbox{repeat s}    & \lstbox{Just [Leq,Any]}\\
\lstbox{innermost s} & \lstbox{Just [Leq,Any]}\\
\hline
\end{tabular}
\end{center}}
\medskip
\caption{\textbf{Exercising compound measures.}}
\label{T:LeqTypes2}
\figskip
\end{figure}


The power of such termination types is illustrated in
\autoref{T:LeqTypes2}. The new type for \hs{full_td}\ shows that we
do not rely on the argument strategy to be non-increasing on the
depth; we may as well use a depth-increasing argument, as long as it
is non-increasing on some constructor count. The new type for
\hs{once_bu}\ is strictly decreasing, and hence, its iteration with
\hs{repeat}\ results in a termination type for \hs{innermost}.
We refer to the paper's online code distribution for details.

\section{Related work}
\label{S:related}


\vspace{-77\in}

We will focus here on related work that deals with or is directly
applicable to programming errors in traversal programming with
traversal strategies or otherwise.


\relatedWorkParagraph{Simplified traversal programming}

In an effort to manage the relative complexity of traversal strategies
or the generic functions of ``Scrap Your Boilerplate'', simplified
forms of traversal programming have been proposed.  The functional
programming-based work of \cite{MitchellR07} describes less generic
traversal schemes (akin to those of \S\ref{S:adviceDefault}), and
thereby suffices with simpler types, and may achieve efficiency of
traversal implementation more easily. The rewriting-based work of
\cite{BrandKV03} follows a different route; it limits programmability
of traversal by providing only a few schemes and few
parameters. Again, such a system may be easier to grasp for the
programmer, and efficiency of traversal implementation may be
achievable more easily. Our work is best understood as an attempt to
uphold the generality or flexibility and relative simplicity (in terms
of language constructs involved) of traversal strategies while using
orthogonal means such as advanced typing or static analysis for
helping with program comprehension.


\relatedWorkParagraph{Implicit strategy extension}

It is fair to say that several challenges relate to strategy
extension. When traversal schemes are not parameterized with generic
functions, as mentioned above, then strategy extension is no longer
needed by the programmer, but expressiveness will be limited.  There
exists a proposal~\cite{DolstraV01} for a language design that
essentially makes strategy extension implicit in an otherwise typeful
setting.  This approach may imply a slightly simpler programming
model, but it is not obvious that it necessarily reduces programming
errors. This question remains open, but we refer to
\cite{Laemmel03-JLAP} for a (dated) discussion of the matter.


\relatedWorkParagraph{Runtime checks}

With respect to our various efforts to declare, infer, and enforce
fallibility properties it is useful to note that Stratego supports the
\emph{with} operator as alternative to the \emph{where} clause for
conditional rules (and as strategy combinator as well), which
indicates that its argument should be a transformation that always
succeeds. When it does not succeed, a run-time exception is raised. It
is reported, anecdotally, for example, by a reviewer of this paper,
that this feature has been helpful for detection of programming
errors. Such annotations can be useful for expressing the intent of
programmers and may be useful both for runtime checking and as input
to static analyses in future systems.


\relatedWorkParagraph{Adaptive programming}

While the aforementioned related work is (transitively) inspired by
traversal strategies \`a la Stratego, there is the independent
traversal programming approach of adaptive programming~\citeAP.
Traversal specifications are more disciplined in this paradigm.
Generally, the focus is more on problem-specific traversal
specifications as opposed to generic traversal schemes.  Also, there
is a separation between traversal specifications and computations or
actions, thereby simplifying some analyses, e.g., a termination
analysis.  The technique of \S\ref{S:adviceReachability} to statically
check for reachable types is inspired by adaptive programming. Certain
categories of programming errors are less of an issue, if any, in
adaptive programming. Interestingly, there are recent efforts to
evolve adaptive programming, within the bounds of functional
object-oriented programming, to a programming paradigm that is appears
to be more similar to strategic programming~\cite{AbdelmegedL07}.


\relatedWorkParagraph{The XML connection}

Arguably, the most widely used kinds of traversal programs in practice
are XML queries and transformations such as those based on
XPath~\cite{XPath}, XSLT~\cite{XSLT}, and XQuery~\cite{XQuery}. Some
limited cross-paradigmatic comparison of traversal strategies and XML
programming has been presented in~\cite{Laemmel07,CunhaV07}. Most
notably, there are related problems in XML programming. For instance,
one would like to know that an XPath query does not necessarily return
the empty node set. The XQuery specification~\cite{XQuery} even
addresses this issue, to some extent, in the XQuery type system. While
XSLT also inherits all XPath-related issues (just as much as XQuery),
there is an additional challenge due to its template mechanism.
Default templates, in particular, provide a kind of traversal
capability. Let us mention related work on analyzing XML programs.
\cite{DongBailey} describes program analysis for XSLT based on an
analysis of the call graph of the templates and the underlying DTD for
the data. In common with our work is a conservative estimation of
sufficient conditions for program termination as well as some form
dead code analysis. There is recent work on logics for XML
\cite{geneves-phd06} to perform static analysis of XML paths and types
\cite{geneves-pldi07}, and a dead code elimination for XQuery programs
\cite{ICSE-XML}.


\relatedWorkParagraph{Properties of traversal programs}

Mentions of algebraic laws and other properties of strategic
primitives and some traversal schemes appear in the literature on
Stratego-like
strategies~\cite{VisserBT98,Essence,Visser03-PHD,BrandKV03,JohannV03,LaemmelPJ03,Laemmel03-JLAP,Reig04,CunhaV07,KaiserL09}.
In \cite{CunhaV07}, laws feed into automated program calculation for
the benefit of optimization (``by specialization'') and reverse
engineering (so that generic programs are obtained from boilerplate
code). In \cite{JohannV03}, specialized laws of applications of
traversal schemes are leveraged to enable fusion-like techniques for
optimizing strategies. We contend that better understanding of
properties of traversal strategies, just by itself, curbs programming
errors, but none of these previous efforts have linked properties to
programming errors. Also, there is no previous effort on performing
static analysis for the derivation of general program properties about
termination, metadata-based reachability, or success and failure
behavior.


\relatedWorkParagraph{Termination analysis}

Our termination analysis is arguably naive in that it focuses on
recursion patterns of traversals. A practical system for a
full-fledged strategic programming language would definitely need to
include existing techniques for termination analysis, as they are
established in the programming languages and rewriting communities. We
mention some recent work in the adjacency of (functional) traversal
strategies.  \cite{GnaedigK08,ThiemannS09} address termination
analysis for rewriting with strategies (but without covering
programmable traversal strategies). \cite{SereniJ05,GieslSST06}
address termination analysis for higher-order functional programs; it
may be possible to extend these systems with awareness for one-layer
traversal and generic functions for traversal. \cite{Abel09} addresses
termination analysis for generic functional programs of the kind of
Generic Haskell~\cite{Hinze00}; the approach is based on type-based
termination and exploits the fact that generic functions are defined
by induction on types, which is not directly the case for traversal
strategies, though.

\section{Concluding remarks}
\label{S:concl}

The ultimate motivation for the  work presented here is to make traversal
programming with strategies easier and safer. To this end, strategy
libraries and the underlying programming languages need to improve so
that contracts of traversal strategies are accurately captured and
statically verified. That is, we envisage that ``design by contract''
is profoundly instantiated for traversal programming with strategies.
These contracts would deal, for example, with (in)fallibility or
measures for termination.

Throughout the paper, we have revealed pitfalls of strategic
programming and discovered related properties of basic strategy
combinators and common library combinators. To respond to these, we have
developed concrete advice on suggested improvements to strategy
libraries and the underlying programming languages:
\begin{itemize}
\item \adviceDefault. \hfill (\S\ref{S:adviceDefault})
\item \adviceFallibility. \hfill (\S\ref{S:adviceFallibility})
\item \adviceCflow. \hfill (\S\ref{S:adviceCflow})
\item \adviceTypecase. \hfill (\S\ref{S:adviceTypecase})
\item \adviceReachability. \hfill (\S\ref{S:adviceReachability})
\item \adviceInferFallibility. \hfill (\S\ref{S:adviceInferFallibility})
\item \adviceProveReachability. \hfill (\S\ref{S:adviceProveReachability})
\item \adviceTermination. \hfill (\S\ref{S:adviceTermination})
\end{itemize}
Such advice comes without any claim of completeness. Also, such advice
must not be confused with a proper design for the ultimate library and
language.

Arguably, some improvements can be achieved by revising strategy
libraries so that available static typing techniques are
leveraged. We demonstrated this path with several Haskell-based
experiments. This path is limited in several respects. First, only
part of the advice can be addressed in this manner. Second, the typing
techniques may not be generally available for languages used in
traversal programming. Third, substantial encoding effort is needed in
several cases.

Hence, our work suggests that type systems of programming languages
need to become more expressive so that all facets of traversal
contracts can be captured at an appropriate level of abstraction and
statically verified in a manner that also accounts for language
usability. We assert that, in fact, strategic programming is in need
of a form of dependent types, an extensible type system, or, indeed,
an extensible language framework that admits pluggable static
analysis.

The development of the paper is based on a series of programming
errors as they arise from a systematic discussion of the process of
design and implementation of strategic programs. Empirical research
would be needed to confirm the relevance of the alleged
pitfalls. However, based on the authors' experience with strategic
programming in research, education, and software development, the
authors can confirm that there is anecdotal evidence for the existence
of the presented problems.

In this paper, we have largely ignored another challenge for strategic
programming, namely \emph{performance}. In fact, disappointing performance
may count as another kind of programming error. Better formal and
pragmatic understanding of traversal programming is needed to execute
traversal strategies more efficiently. Hence, performance is suggested
as a major theme for future work. There is relevant, previous work on
fusion-like techniques for traversal strategies \cite{JohannV03},
calculational techniques for the transformation of traversal
strategies \cite{CunhaV07}, and complementary ideas from the field of
adaptive programming \cite{LieberherrPO04}.


\bigskip

{\footnotesize

\noindent
\textbf{Acknowledgments}. Simon Thompson has received support by the
Vrije Universiteit, Amsterdam for a related research visit in
2004. The authors received helpful feedback from the LDTA reviewers
and the discussion at the workshop. Thanks are due to the reviewers of
the initial journal submission who provided substantial advice. Part
of the material has been used in Ralf L\"ammel's invited talk at
LOPSTR/PPDP 2009, and all feedback is gratefully acknowledged. Much of
the presented insights draw from past collaboration and discussions
with Simon Peyton Jones, Karl Lieberherr, Claus Reinke, Eelco Visser,
Joost Visser, and Victor Winter.

}


{

\noskip

\bibliography{paper}
\bibliographystyle{abbrv}

}

\end{document}